\documentclass[twocolumn]{aastex631}
\newcommand{\msun}{M_{\odot}}
\usepackage[version=4]{mhchem}
\usepackage{amsmath,amssymb,amsthm,mathtools,mathdots,nicefrac,tikz,bbding,stmaryrd,tikz-cd,cases,mathrsfs,soul,verbatim, dsfont, upgreek}

\usepackage{breqn}
\usepackage{amsmath}
\usepackage{ulem}
\usepackage[colorinlistoftodos]{todonotes}

\newcommand{\kms}{$\rm km~s^{-1}$\,}

\newcommand{\HtO }{$\rm H_2O$}

\graphicspath{{./}{figures/}}

\shorttitle{Orion source I: Accretion and Outflow}
\shortauthors{Wright et al.}
\begin{document}
\title{Accretion and Outflow in Orion-KL Source I}

\correspondingauthor{Melvyn Wright}
\email{melvyn@berkeley.edu}

\author{Melvyn Wright}
\affiliation{Department of Astronomy, University of California, 501 Campbell Hall, Berkeley CA 94720-3441, USA}

\author{Brett A. McGuire}
\affil{Department of Chemistry, Massachusetts Institute of Technology, Cambridge, MA 02139, USA}
\affil{National Radio Astronomy Observatory, Charlottesville, VA 22903, USA. 
ORCiD: 0000-0003-1254-4817}

\author[0000-0001-6431-9633]{Adam Ginsburg}
\affil{Department of Astronomy, University of Florida
211 Bryant Space Science Center
P.O Box 112055, Gainesville, FL 32611-2055 USA}

\author{Tomoya Hirota}
\affiliation{Mizusawa VLBI Observatory, National Astronomical Observatory of Japan, 
2-12, Hoshigaoka, Mizusawa, Oshu, Iwate 023-0861, Japan}
\affiliation{The Graduate University for Advanced Studies, SOKENDAI, 2-21-1 Osawa, Mitaka, Tokyo 181-8588, Japan}

\author{John Bally}
\affil{CASA, University of Colorado, 389-UCB, Boulder, CO 80309, USA}

\author{Ryan Hwangbo }
\affiliation{Department of Astronomy, University of California, 501 Campbell Hall, Berkeley CA 94720-3441, USA}

\author{T. Dex Bhadra}
\affiliation{Department of Astronomy, University of California, 501 Campbell Hall, Berkeley CA 94720-3441, USA}

\author{Chris John}
\affiliation{Troy High School,
2200 Dorothy Ln, Fullerton, CA 92831, USA}
\affiliation{Department of Astronomy, University of California, 501 Campbell Hall, Berkeley CA 94720-3441, USA}

\author{Rishabh Dave}
\affiliation{Department of Astronomy, University of California, 501 Campbell Hall, Berkeley CA 94720-3441, USA}


\begin{abstract}

We present ALMA observations of SiO, SiS, \HtO , NaCl, and SO line emission at $\sim$30 to 50 mas resolution. These images map the molecular outflow and disk of Orion Source I (SrcI) on $\sim$12 to 20 AU scales.
Our observations show that the flow of material around SrcI creates a turbulent boundary layer in the outflow from SrcI which may dissipate angular momentum in the rotating molecular outflow into the surrounding medium.
Additionally, the data suggests that the proper motion of SrcI may have a significant effect on the structure and evolution of SrcI and its molecular outflow. As the motion of SrcI funnels material between the disk and the outflow, some material may be entrained into the outflow and accrete onto the disk, creating shocks which excite the NaCl close to the disk surface.

\end{abstract}

\section{Introduction} \label{sec:intro}

Massive stars play an important role in numerous astrophysical processes, but their formation and evolution remains ambiguous. Current theory regarding high-mass ($>8 M_\odot$) protostellar evolution strongly suggests that accretion onto the central parts via disks, and the specific geometry and rate are key variables that dictate the characteristics of the resulting star \citep{Caretti_o_Garatti_2017, Hosokawa2010}. Efforts to find a mechanism that reconciles the in-falling accreted material and outgoing outflows/radiation have resulted in numerous theories regarding the necessary rate and three-dimensional accretion flow geometry e.g., 
\citep{Hosokawa2010, Whitney_2013, Kuiper_2011}.
Other theories, such as stellar mergers  \citep{Bonnell_1998} and competitive accretion  \citep{Bonnell_2008} have also been considered. However, regardless of the specifics, the mechanism must be able to move material into the protostar for it to continue growing. 

Study of the evolution of high-mass protostars is possible in the Kleinmann-Low Nebula in Orion (Orion-KL). At a distance of 415pc \citep{Menten2007, Kim2008,Kounkel2018}, Orion-KL is the nearest interstellar cloud with observable massive star formation. 

Observations of Orion-KL and simulations suggest that three stars were involved in a dynamical decay event approximately 550 years ago, which ejected them from a multi-star system \citep{Bally2017, Bally2020}. 
One of these stars, the Becklin-Neugebauer (BN) object, has a proper motion of $\sim$26 \kms 
northwestward. Radio Source I (SrcI) is another star involved, and moves at $\sim$10 \kms southward. Finally, near-infrared source x (SrcX) moves at $\sim$55 \kms southeastward \citep{Dzib_2017,Luhman_2017,Rodgriguez2017}. 

Direct observation of the central object in SrcI is not possible as the high opacity of the continuum emission from the SrcI disk conceals the central source.
The specific nature and mass of this object has been the subject of extensive study and theorization. Estimation-through conservation of linear momentum and dynamical simulation best agreed with SrcI being a $\sim20 M_\odot$ binary (proto-)star system \citep{Rodriguez2017, Rodriguez2020, Bally2020,Goddi2011b}.
\cite{Ginsburg2018} found SrcI to have a mass of $15 \pm 2 M_\odot$ by fitting the rotation curve of the \ce{H2O} $5_{5,0}-6_{4,3}$ spectral line.

To build stronger constrains on its nature, we explore the outflow SrcI produces due to its close association with accretion. Outflows provide an avenue to carry excess angular momentum out of the system, enabling accreting matter to fall upon the star. This transfer of momentum and kinetic energy influences star formation in surrounding regions.

SrcI has been well-studied as it is the closest observable high-mass protostar with a circumstellar disk and molecular outflow \citep{Hirota2014, Plambeck2016,Ginsburg2018}. The molecular outflow from SrcI is prominent in SiO, SO, and SiS \citep{Hirota2020,Wright2020,Lopez-Vazquez2020}. The disk and outflow are associated with SiO and \ce{H2O} masers \citep{Reid2007,Goddi2009,Goddi2011,Plambeck2009,Matthews2010, Niederhofer2012,Greenhill2013}. The mechanisms that drive outflows, such as the entrainment of molecular gas by a higher-velocity jet or an MHD-driven outflow by magneto-centrifugal disk winds, are still poorly understood \citep{Commercon2022}.

In this paper, we re-examine our observational data from ALMA, using several molecular spectral lines that trace key structures in SrcI to look for signatures of accretion in SrcI.

\section{Observations} \label{sec:data}

We used data at $223$ GHz (ALMA Band 6; B6) and $340$ GHz (ALMA Band 7; B7) as described in \citet{Ginsburg2018}. This data was processed into images using both uniform and natural weighting of the uv-data. The Brigg's visibility weighting was set to robust $=0$ to avoid the excessive weight being given to visibilities in relatively sparsely filled regions of the uv-plane by uniform weighting. For naturally-weighted images, we set robust $=2$ to map the low-brightness, more extended emission. The observational parameters are given in Table 1 for the uniformly-weighted uv-data.

Within these bands, several molecular species have been identified as present in SrcI: SiO, SiS, \ce{H2O}, NaCl, and SO. Based on the shape and location of the emission from these lines, SiO and SiS have been interpreted as tracers of a wide-angle outflow along the minor axis of a circumbinary disk, while \ce{H2O} and NaCl have a strong presence on the surface of/within the circumbinary disk. The structure of SO is more complex and will be further explored in later sections.

The high opacity at $220$ - $330$ GHz of the continuum emission from the SrcI disk conceals the internal structure of the disk. The less opaque $99$ GHz continuum observations from \cite{Wright2022} will also be analyzed as they reveal details of disk's structure. Emission from molecules associated with the disk gives us information about the structure and kinematics of the SrcI disk and its interaction with the material in its immediate environment.

Hereafter, we discuss the specific molecular lines $^{28}$SiO,
%$\quad v=0\quad(J=5-4)$, 
$^{29}$SiO,
%$\quad v=0 \quad (J=5-4)$, 
SiS,
%$\quad v=0 \quad (J=12-11)$ , 
\ce{H2O}, NaCl, and SO to analyze the structure and kinematics of molecular lines closely associated with SrcI for evidence of accretion.
The spectral lines discussed in this paper are listed in Table 2.

\section{Results} \label{sec:results}

In this section we present the spectral line data as the average
intensity and average velocity of the emission across the spectral line. These were derived from the intensity and velocity of each spectral channel. The contours map the intensity of the emission. This may be expressed in units of Jy/beam or Kelvin.
We give the equivalent Rayleigh-Jeans brightness in Kelvin for the peak value in the image. The average velocity is shown as a color scale in units of \kms.

\subsection{The Molecular Outflow; SiO and SiS}

%{\color{blue} FIGURE 1}

\autoref{fig:sio_mom0+1} shows the $217.105$ GHz $^{28}$SiO$\quad v=0\quad(J=5-4)$ emission with a peak brightness of $1078$ K at a velocity resolution of $2$ \kms. This figure also shows $214.385$ GHz $^{29}$SiO$\quad v=0\quad(J=5-4)$ emission with a peak brightness of $18300$ K at a velocity resolution of $2$ \kms. The high peak brightness temperature of the $^{29}$SiO emission indicates the presence of masers. 
 
\autoref{fig:sis_mom0+1} shows the $217.818$ GHz SiS$\quad v=0\quad(J=12-11)$ emission with a peak brightness of $409$ K with a velocity resolution of $2$ \kms. The SiS emission traces a shell-like structure along the boundaries of the SiO outflow \citep{Wright2020}. This higher-resolution SiS image reveals the complex structure it traces.

\subsection{SO}

\autoref{fig:son344GHz} shows the 344.31 GHz SO emission with natural weighting of the u-v data to emphasize the larger scale structure. The peak brightness with 1 \kms velocity resolution is 375 K.

\subsection{\ce{H2O} and NaCl Emission}

\autoref{fig:nacl_mom0+1} shows the \ce{H2O} and \ce{NaCl} Emission. The 232.687 GHz \ce{H2O} line (5(5,0), v$_{2}$=1 - 6(4,3), v$_{2}$=1) is more closely associated with the SrcI disk than the SiO and SiS, which primarily trace the molecular outflow.

\autoref{fig:nacl_mom0+1} also shows the 232.51 GHz NaCl$\quad v=1\quad(J=18-17)$ emission with a peak brightness of $197$ K with a velocity resolution of $2$ \kms.

\section{Discussion} \label{sec:discussion}

In this section we discuss each of the molecular lines observed, and the regions of the disk and outflow that each traces.
The distribution of the molecular lines depends on the
formation and destruction of the molecules, and we include a discussion of the chemical interactions between the outflow and ambient material. The structure and kinematics of the SiS suggest that ambient material may be entrained into the molecular outflow
leading to the export of angular momentum from the outflow. We discuss the evidence for Bondi-Hoyle-Lyttleton accretion onto the disk around Source I, and the evolution of the disk and the protostar(s)

\subsection{The Bipolar Outflow}

\citet{Hirota2017} fitted a hollow cone model to the 484.056 GHz Si$^{18}$O $J=12-11$ line and constrained the launching radius and outward velocity of the outflow to be $>$10 AU and $\sim$10 km s$^{-1}$, respectively. From these results, \citet{Hirota2017} concluded that the rotating outflow in SrcI is likely driven by a magneto-centrifugal disk wind. 
\cite{Hirota2017} note that the critical density required to excite the $J=12-11$ transition ($6\times10^7$ cm$^{-3}$) is higher than those of $J=2-1$ at 86 GHz ($3\times10^5$ cm$^{-3}$) and $J=5-4$ at 217 GHz ($4\times10^6$ cm$^{-3}$).  Additionally, the $^{28}$SiO molecular line may be highly opaque, as discussed by \citet{Lopez-Vazquez2020}. The  Si$^{18}$O isotopologue is expected to be optically thin, as it traces regions of higher density than the optically thick Si$^{16}$O lines.

\begin{figure}
% trim left bottom right top
%\includegraphics[width=1.0\columnwidth, clip, trim=3cm 3.0cm 2.0cm 1cm] {sio_mom0+1.pdf}
%\includegraphics[width=1.0\columnwidth, clip, trim=3cm 3.7cm 2.0cm 1cm]{29sio_mom0+1.pdf}
\includegraphics[width=0.5\textwidth, clip, trim=3cm 3cm 2.0cm 1cm]{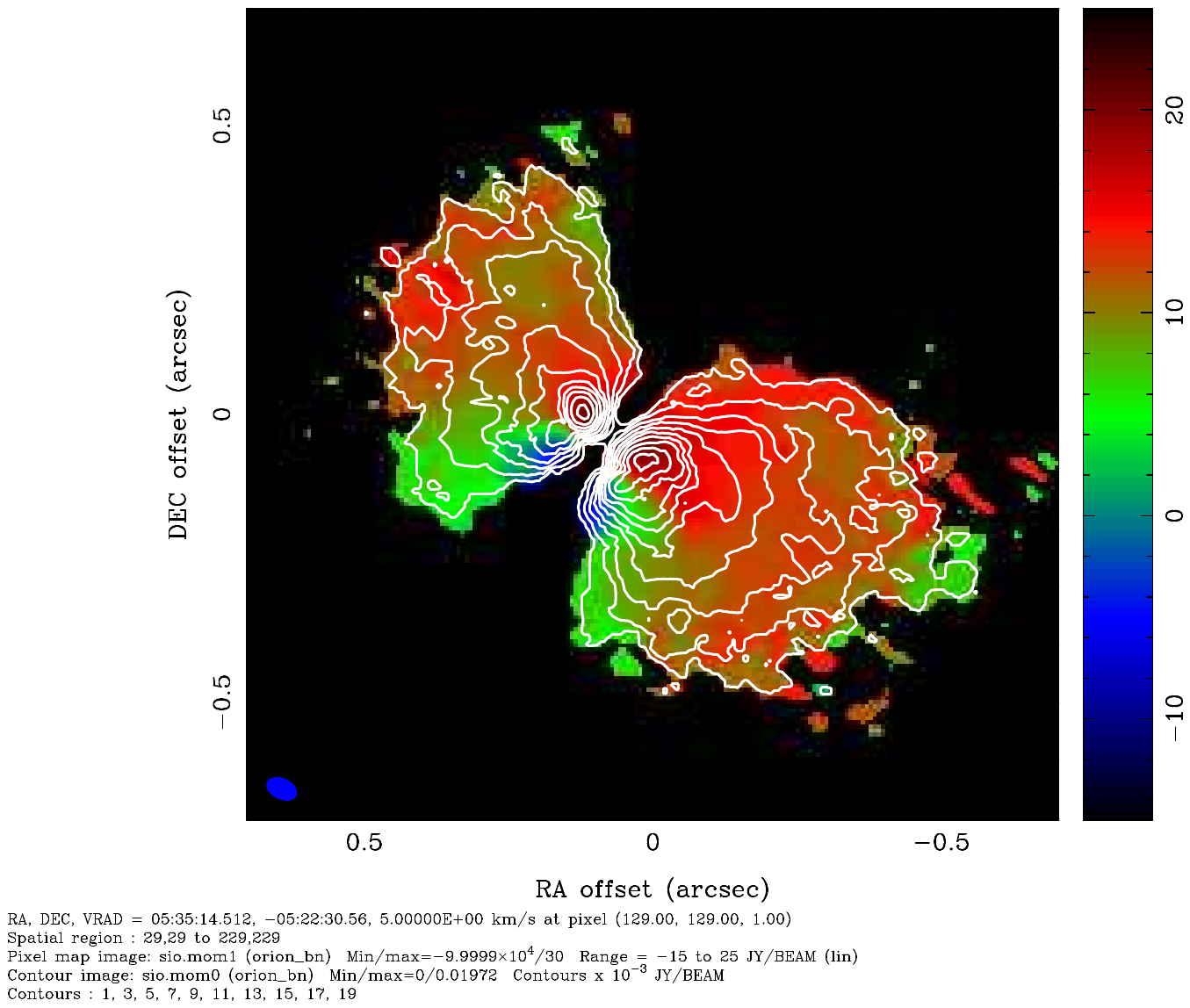}
\includegraphics[width=0.5\textwidth, clip, trim=3cm 3.6cm 2.0cm 1cm]{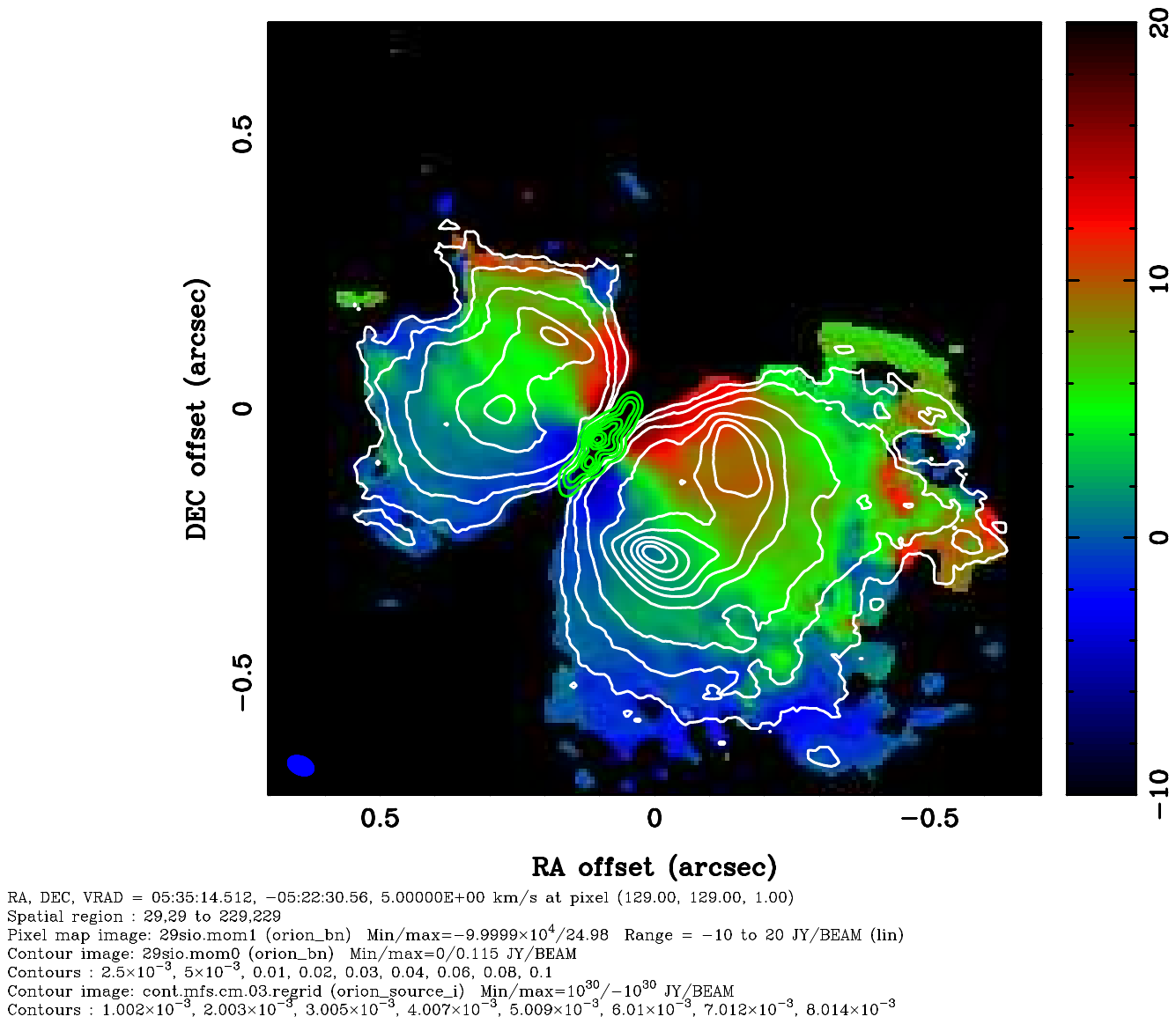}
\caption{
{\bf{ SiO and  $^{29}$SiO emission from -20 to +30 km s$^{-1}$.}}
\textbf{(Left)} 
The color image shows the moment 1 velocity image for the 217.10 GHz SiO $v=0$ $J=5-4$ with velocities from -15 to +25 \kms indicated in the color wedge. The peak brightness with 2 km s$^{-1}$ velocity resolution is 1078 K.
The white contours show the SiO emission, integrated over -20 to +30 km s$^{-1}$, at 1, 3, 5, 7, 9, 11, 13, 15, 17, 19  mJy/beam. 
\textbf{(Right)}
The color image shows the moment 1 velocity image for the 214.39 GHz $^{29}$SiO $v=0$ $J=5-4$ emission with velocities from -10 to +20 \kms indicated in the color wedge. The peak brightness with 2 km/s velocity resolution is 18300 K, indicating maser emission.
The white contours show the $^{29}$SiO emission, integrated over -20 to +30 km s$^{-1}$, at 2.5, 5, 10, 20, 30, 40, 60, 80, 100 mJy/beam. 
The green contours near the center show the 99GHz continuum emission with contours at 1, 2, 3, 4, 5, 6, 7, 8 mJy/beam.
Both SiO isotopes trace the wide-angle outflow along the minor axis of the disk, with a clear velocity gradient along the major
axis close to the disk. Further from the disk, the velocity pattern has a more complex structure.
The synthesized beam for the SiO emission (FWHM 54$\times$34 mas, PA 65\degr) is indicated in blue in the lower left. 
 The continuum data has been convolved by a 30 mas FWHM beam.
}
\label{fig:sio_mom0+1}
\end{figure}

\begin{figure}
% trim left bottom right top
\includegraphics[width=1.0\textwidth, clip, trim=3cm 3.7cm 2.5cm 1cm]{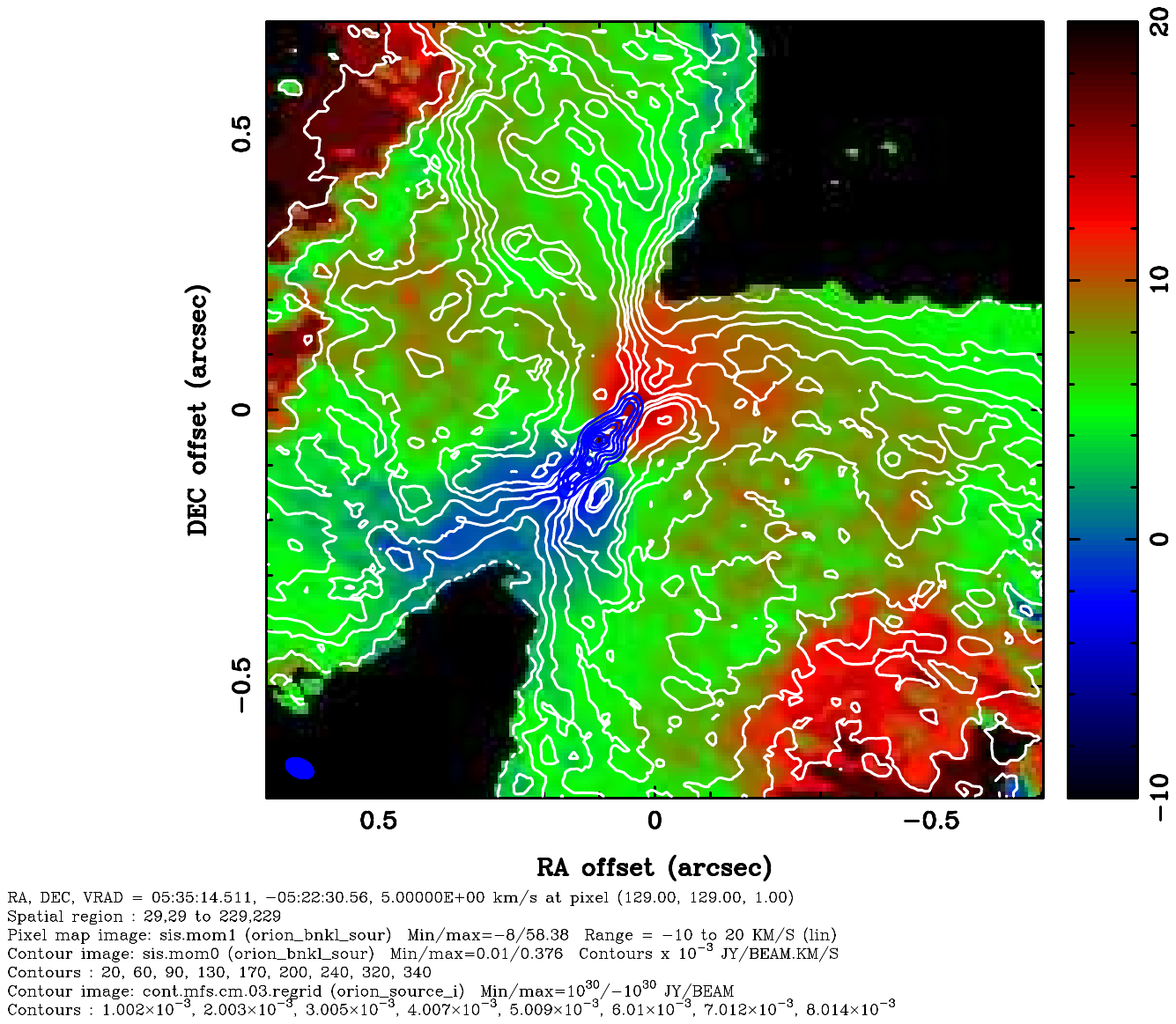}
\caption{
 {\bf{SiS emission from -10 to +20 km s$^{-1}$.}}
The color image shows the moment 1 velocity image for the 217.82 GHz SiS $v=0$ $J=12-11$ emission with velocities from -10 to +20 \kms indicated in the color wedge. The peak brightness with 2 km/s velocity resolution is 409 K.
The white contours show the SiS emission, integrated over -20 to +30 km s$^{-1}$, with contours at 20, 60, 90, 130, 170, 200, 240, 320, 340 mJy/beam $\times$ \kms.
The blue contours show the 99GHz continuum emission with contours at 1, 2, 3, 4, 5, 6, 7, 8 mJy/beam.
There is a clear velocity gradient along the major
axis close to the disk. Further from the disk, SiS has a more complex spacial and velocity structure.
The synthesized beam for the SiS emission, (FWHM 54$\times$34 mas, PA 66\degr), is indicated in blue in the lower left.The continuum data has been convolved by a 30 mas FWHM beam.
\label{fig:sis_mom0+1}}
\end{figure}

\begin{figure}
% trim left bottom right top
\includegraphics[width=1.0\columnwidth, clip, trim=3cm 4.0cm 2.5cm 1cm]{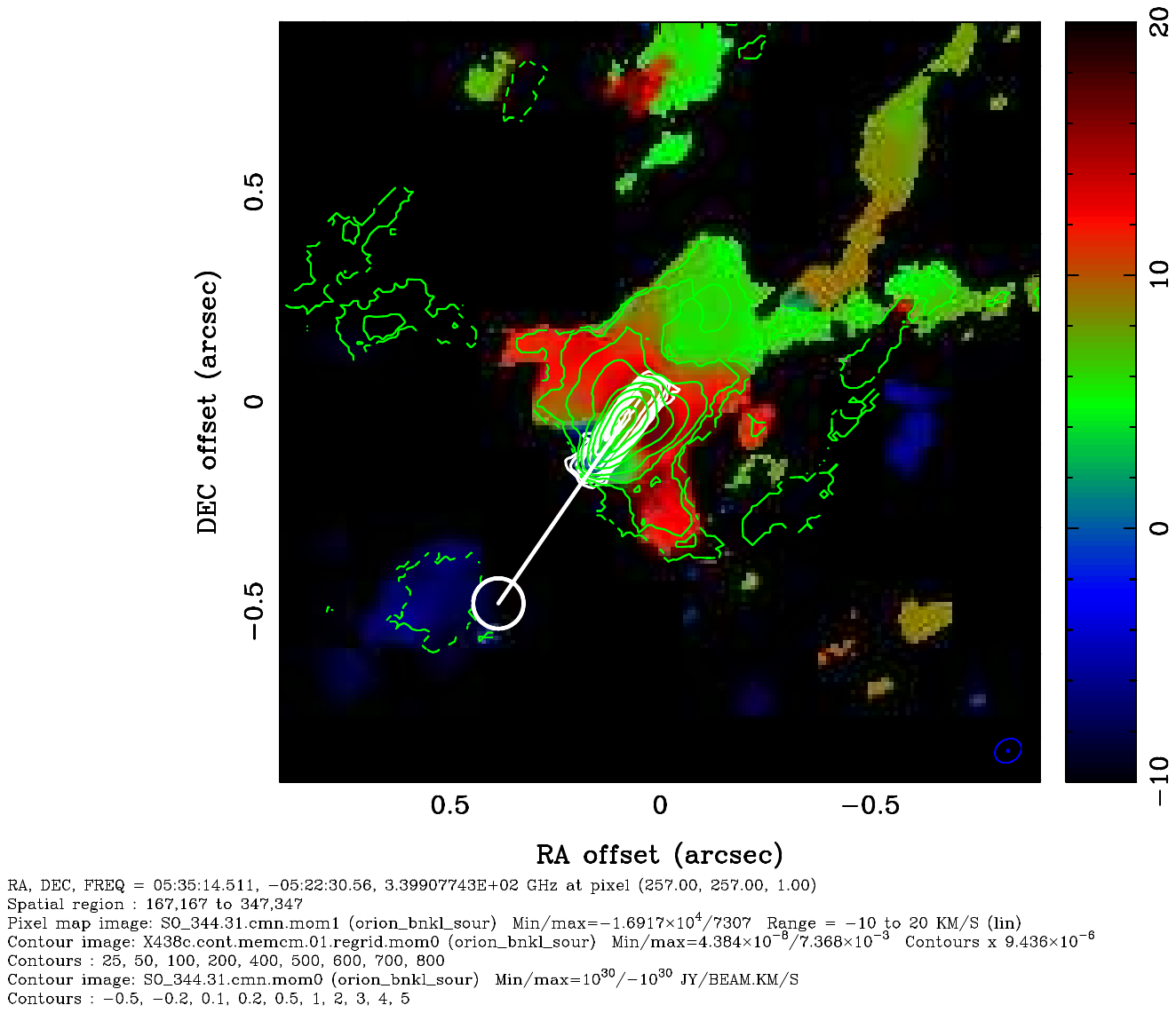}
\caption{
{\bf SO 344.31 GHz emission with natural weighting of the u-v data to emphasize the larger scale structure.}
The color image shows the moment 1 velocity image with velocities from -10 to +20 \kms indicated in the color wedge.
 The peak brightness with 1 km/s velocity resolution is 375 K.
Green contours show the SO emission integrated over a velocity range -20 to +30 km/s. Contour levels are at -0.5, -0.2, 0.1, 0.2, 0.5, 1.0, 2.0, 3.0, 4.0, 5.0 Jy/beam. 
The SO line shows an envelope of emission around the disk, with a steep gradient at the SE end of the disk, and an extended tail to the NW of the disk. There are also emission features to the NE and SW which trace the outflow from the disk as seen in
SiO and SiS emission.
The synthesized beam  FWHM (66$\times$54 mas, PA -56\degr) is indicated in blue in the lower right.
White contours show the 340 GHz continuum emission from the disk at 10 mas resolution. Contour levels are at 25, 50, 100, 200, 400, 500, 600, 700, 800 K.
The white vector shows the proper motion of SrcI in 100 yr.
The open circle radius shows the RMS error in 100 years
\citep{Rodgriguez2017}.
}
%Misalignment of the SrcI disk with its proper motion may produce the turbulent wake seen in SO emission.}
\label{fig:son344GHz}
\end{figure}

\begin{figure}
% trim left bottom right top
\includegraphics[width=0.5\textwidth, clip, trim=3cm 3.7cm 1.3cm 1cm]{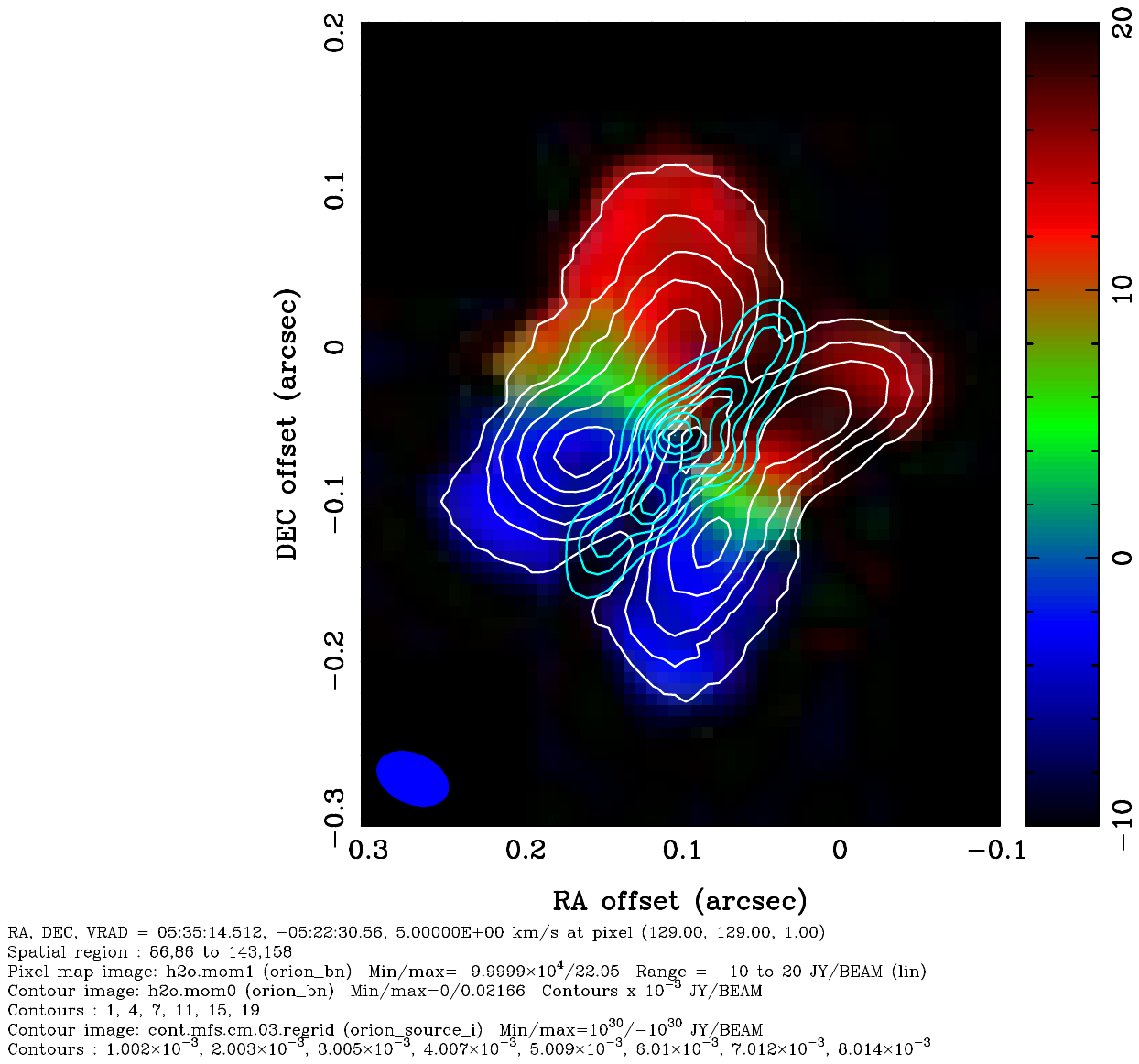}
\includegraphics[width=0.55\textwidth, clip, trim=1cm 3.7cm 1.3cm 1cm]{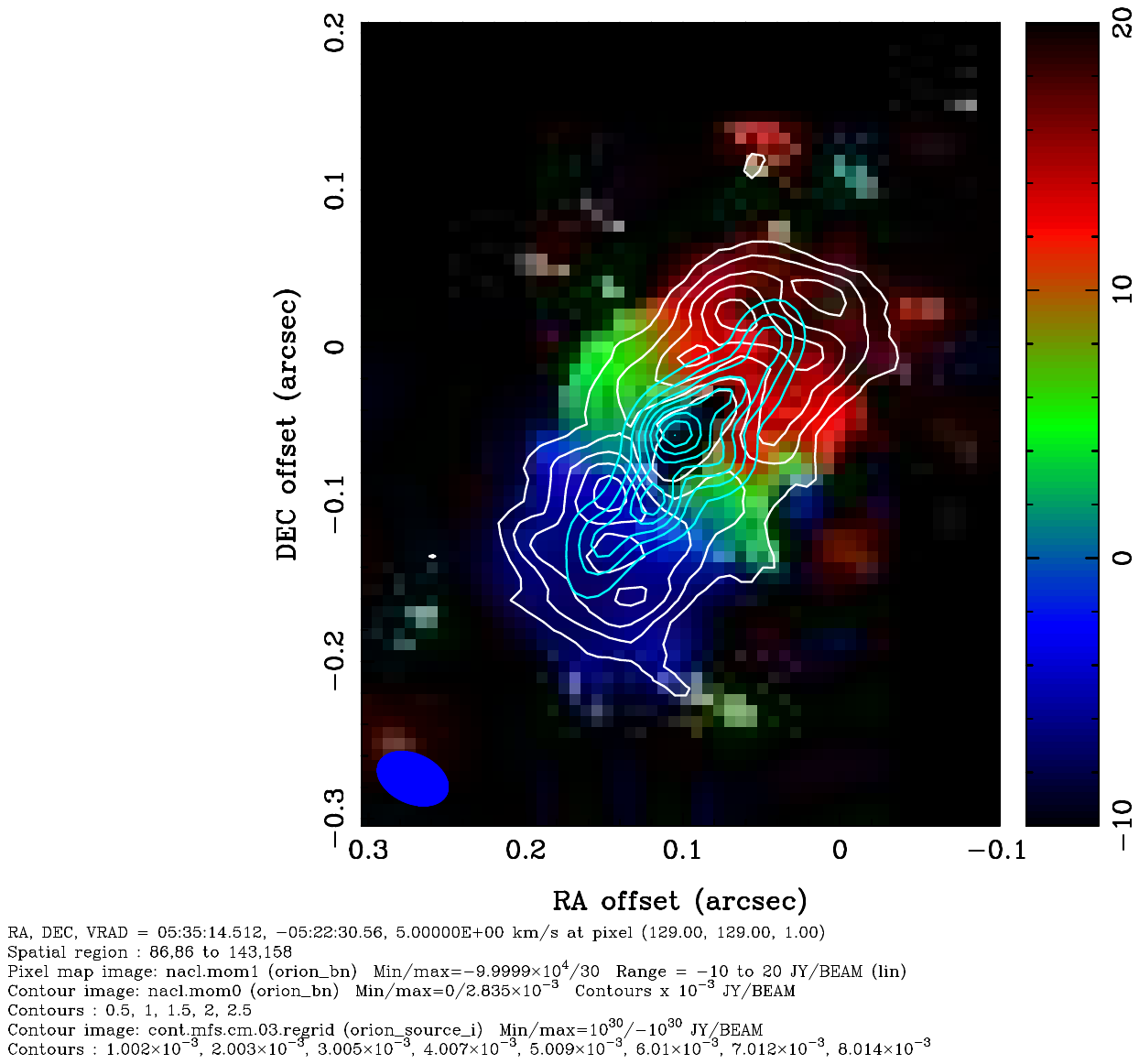}
\caption{
{\bf{ \ce{H_2O} and NaCl emission from -10 to +20 km s$^{-1}$. }}
\textbf{(Left)}
The color image shows the moment 1 velocity image for the 232.69 GHz \ce{H_2O} emission with velocities from -10 to +20 \kms indicated in the color wedge. The peak brightness with 2 km/s velocity resolution is 1200 K.
The white contours map the moment 0 velocity image of the \ce{H_2O} emission integrated over -20 to +30 \kms at contour levels of 1, 4, 7, 11, 15, and 19 mJy/beam.
\textbf{(Right)}
The color image shows the moment 1 velocity image for 232.51 GHz NaCl $v=1$ $J=18-17$ emission with velocities from -10 to +20 \kms indicated in the color wedge. The peak brightness with 2 km/s velocity resolution is 196 K.
The white contours show the NaCl emission, integrated over -20 to +30 km s$^{-1}$, at 1, 4, 7, 11, 15, 19 mJy/beam.
The blue contours show the 99GHz continuum emission  with contours at 1, 2, 3, 4, 5, 6, 7, 8 mJy/beam.
Both NaCl, and  \ce{H2O} show a strong velocity gradient along the disk major axis. 
The NaCl emission is more closely associated with the continuum disk and extends to greater radii along the disk major axis than the \ce{H2O} emission.
The synthesized beam for the  NaCl, and  \ce{H2O} (FWHM 47$\times$31 mas, PA 66 \degr) is indicated in blue in the lower left. The continuum data has been convolved by a 30 mas FWHM beam.
\label{fig:nacl_mom0+1}}
\end{figure}

\begin{figure*}
% trim left bottom right top
\includegraphics[width=1.0\columnwidth, clip, trim=3cm 4.3cm 2.7cm 1cm]{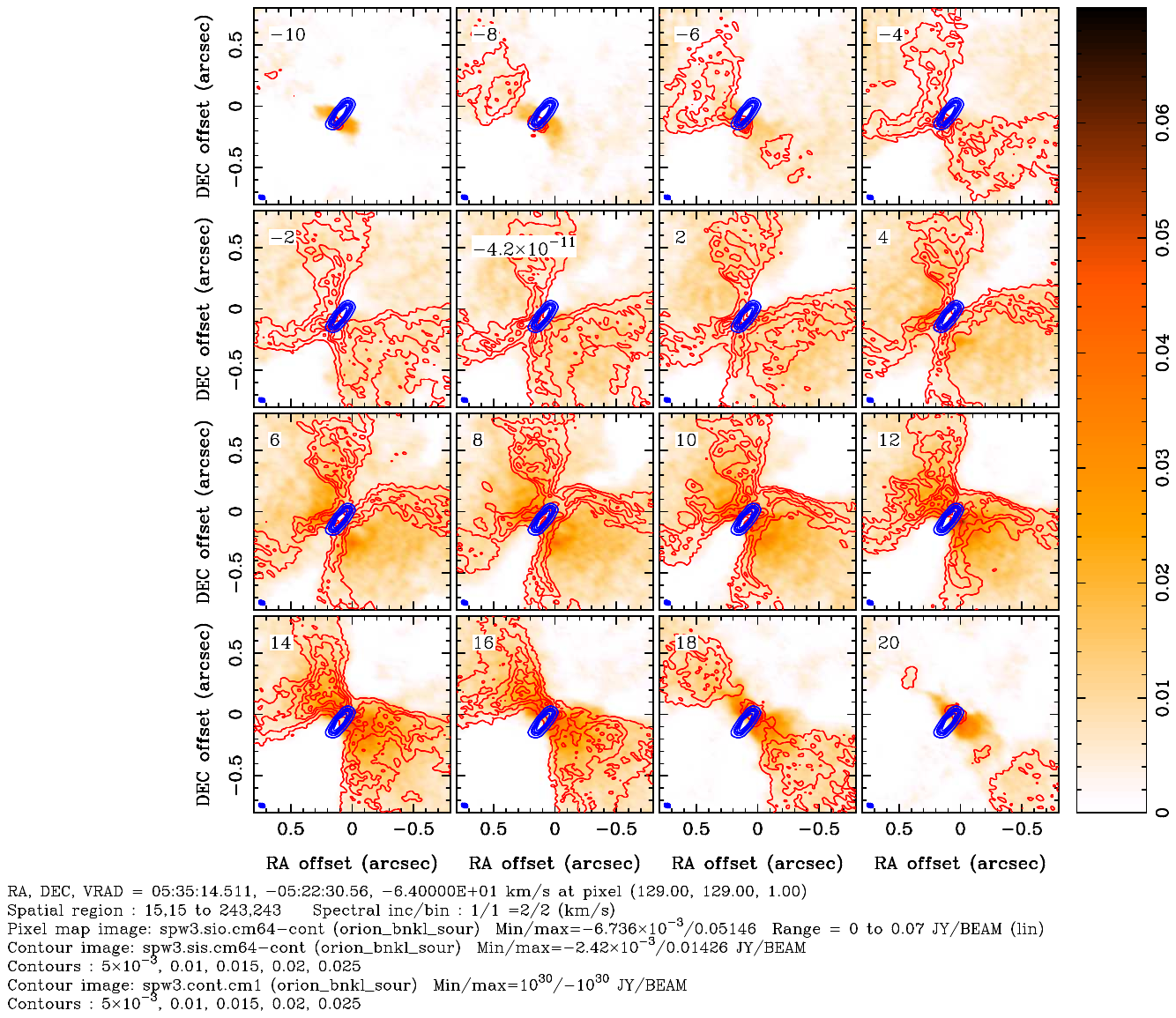}
\caption{
{\bf{Outflow from the SrcI disk mapped in SiO and SiS emission.}}
The color image shows SiO (217.10498 GHz) in 2 \kms channels
with brightness from 0 to 0.07 Jy/beam indicated in the color wedge. LSR velocities in \kms are shown in the upper left.
Red contours show SiS 217.81766 GHz emission in 2~\kms~channels, with contour levels of 5 10 15 20 25 mJy/beam.
Blue contours show 218 GHz continuum with contour levels of 5 10 15 20 25 mJy/beam.
The SiS emission traces a shell-like structure along the boundaries of the SiO outflow .
 The synthesized beam FWHM 54$\times$34 in PA 65\degr is indicated in blue in the lower left.
\label{fig:sio+sis}}
\end{figure*}

Both SiO isotopes shown here in \autoref{fig:sio_mom0+1} clearly trace the wide-angle outflow in SrcI, with the velocity structure growing in complexity away from the disk. In \autoref{fig:sio_mom0+1}, a clear velocity gradient across the (projected) minor axis of the disk can be seen in regions close to the central source. The smooth blue-green-red transition from SE-NW indicates that the rotational velocity of these molecules dominates in these regions. Further away from the center, the smooth velocity gradient disappears and the velocities become "mixed." The $^{28}$SiO image shows little blue-shifted emission beyond the small amount present close the center, with the outer regions of the outflow dominated by an irregular patchwork of green and red-shifted emission. A similar pattern can be seen in the $^{29}$SiO. These features are roughly consistent with the rotating expanding model presented by \citet{Hirota2017}.
\autoref{fig:sio+sis} shows the velocity channel maps of the outflow seen in SiO and SiS emission. At 54$\times$34 mas 
(22$\times$14 au) resolution, the SiS image shows a shell-like structure. The visible velocity structure suggests that there is a turbulent layer on the outside of the SiO emission.

\begin{figure*}
% trim left bottom right top
\includegraphics[width=0.5\textwidth, clip, trim=1cm 1cm 8.5cm 2cm]{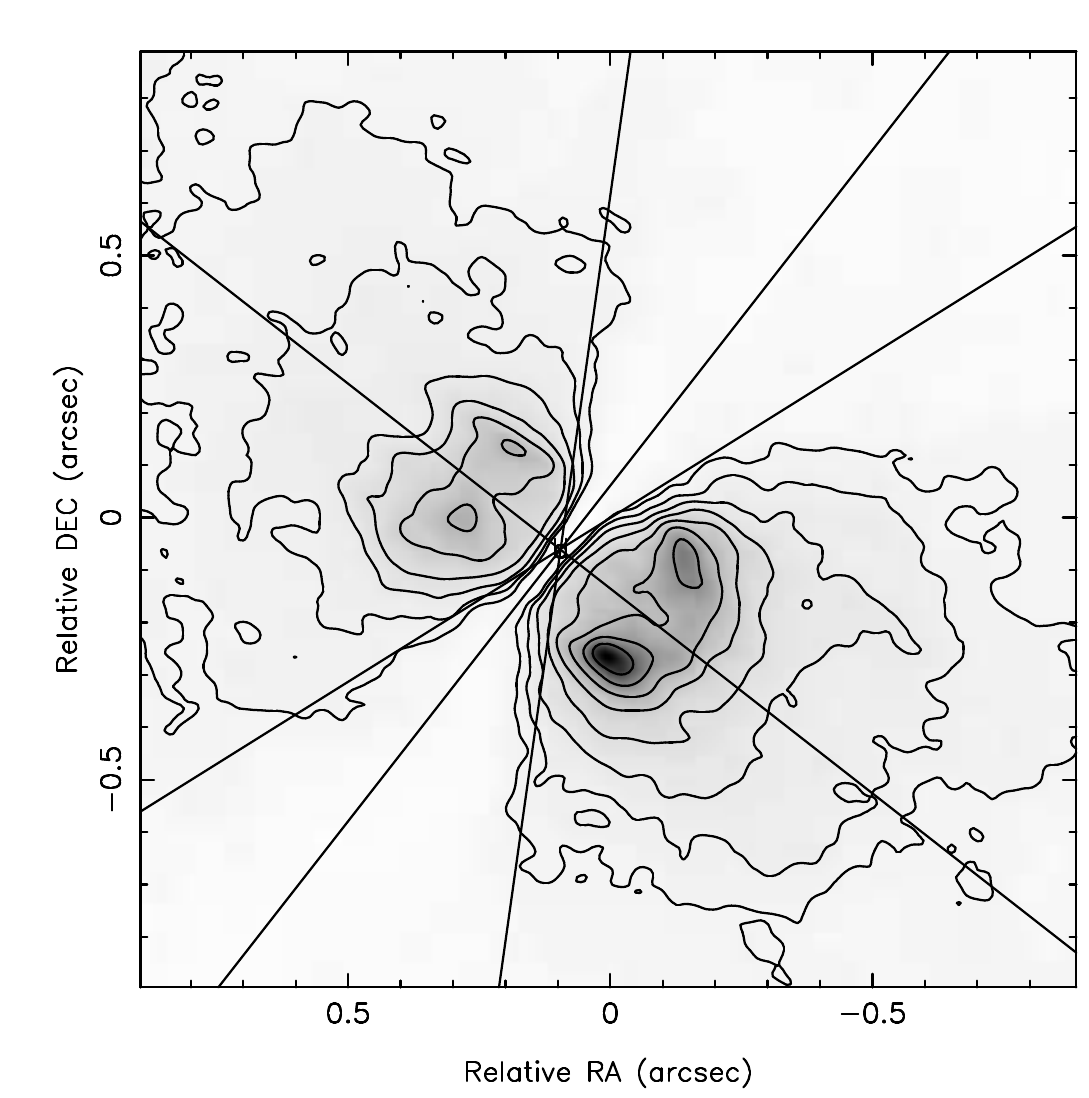}
\includegraphics[width=0.5\textwidth, clip, trim=1cm 1cm 8.5cm 2cm]{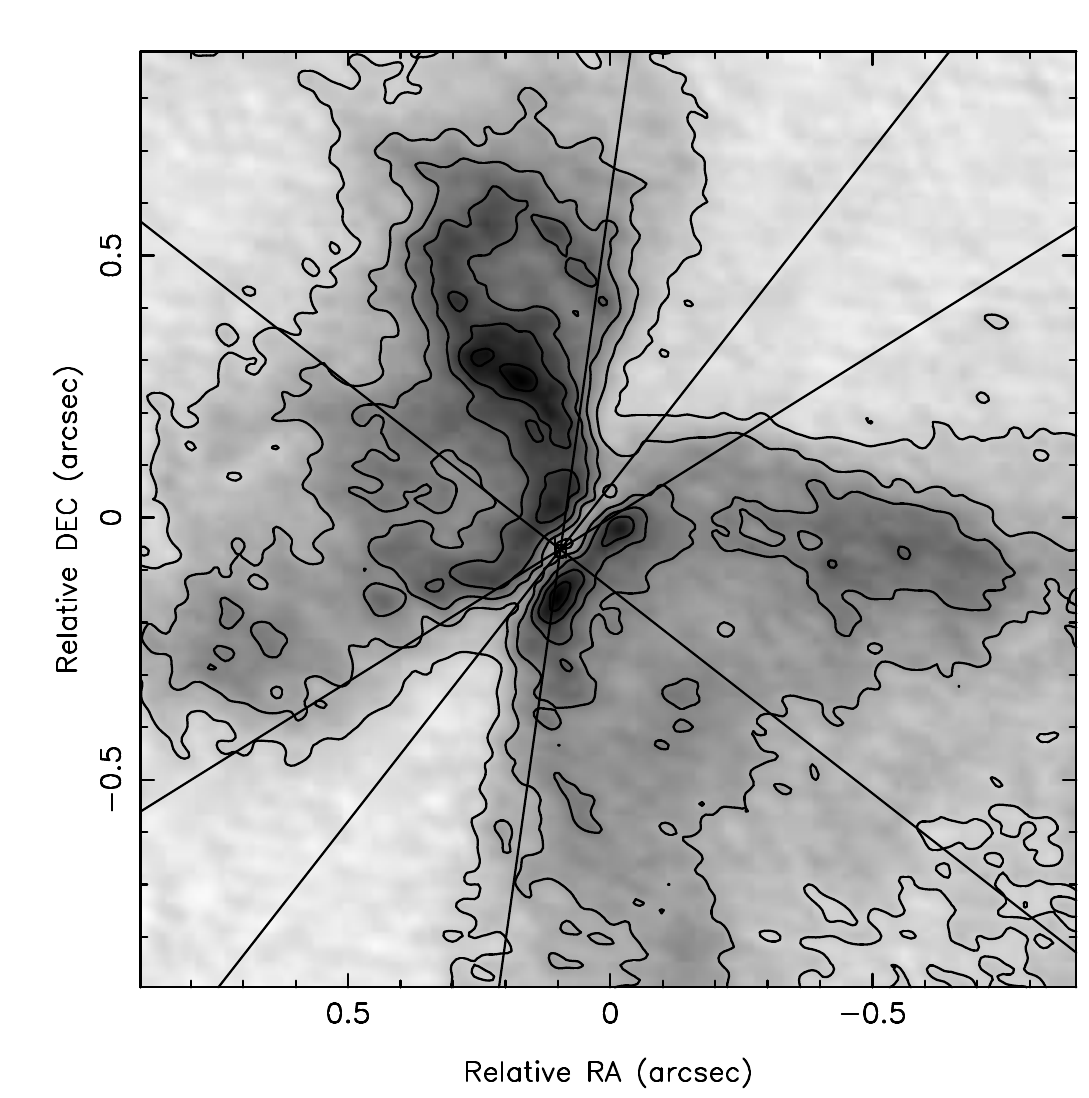}
\caption{{\bf{\ce{^{29}SiO}, and SiS emission integrated over -10 to +20 km s$^{-1}$.}} 
 The 4 lines are drawn along the outflow edges in position angles 122 and 172\degr, along the circumbinary disk major axis in position angle 142\degr, and along the outflow axis in position angle 52\degr.
 \textbf{Left:} \ce{^{29}SiO} 214.39 GHz. 
 \textbf{Right:} SiS 217.82 GHz.
\label{fig:29sio.4cuts}}
\end{figure*}

\autoref{fig:29sio.4cuts} shows the 214.39 GHz \ce{^29SiO} $v=0$ $J=5-4$, and 217.82 GHz SiS $v=0$ $J=12-11$ emission integrated over -10 to +20 km s$^{-1}$. Four cuts are drawn across the outflow: along the outflow edges (PA = 122$\degr$ \& 172$\degr$), along the circumbinary disk major axis (PA = 142\degr), and along the outflow axis (PA = 52\degr).
 
\begin{figure*}
% trim left bottom right top
\includegraphics[width=1.0\columnwidth, clip, trim=1cm 1.1cm 1.8cm 1cm]{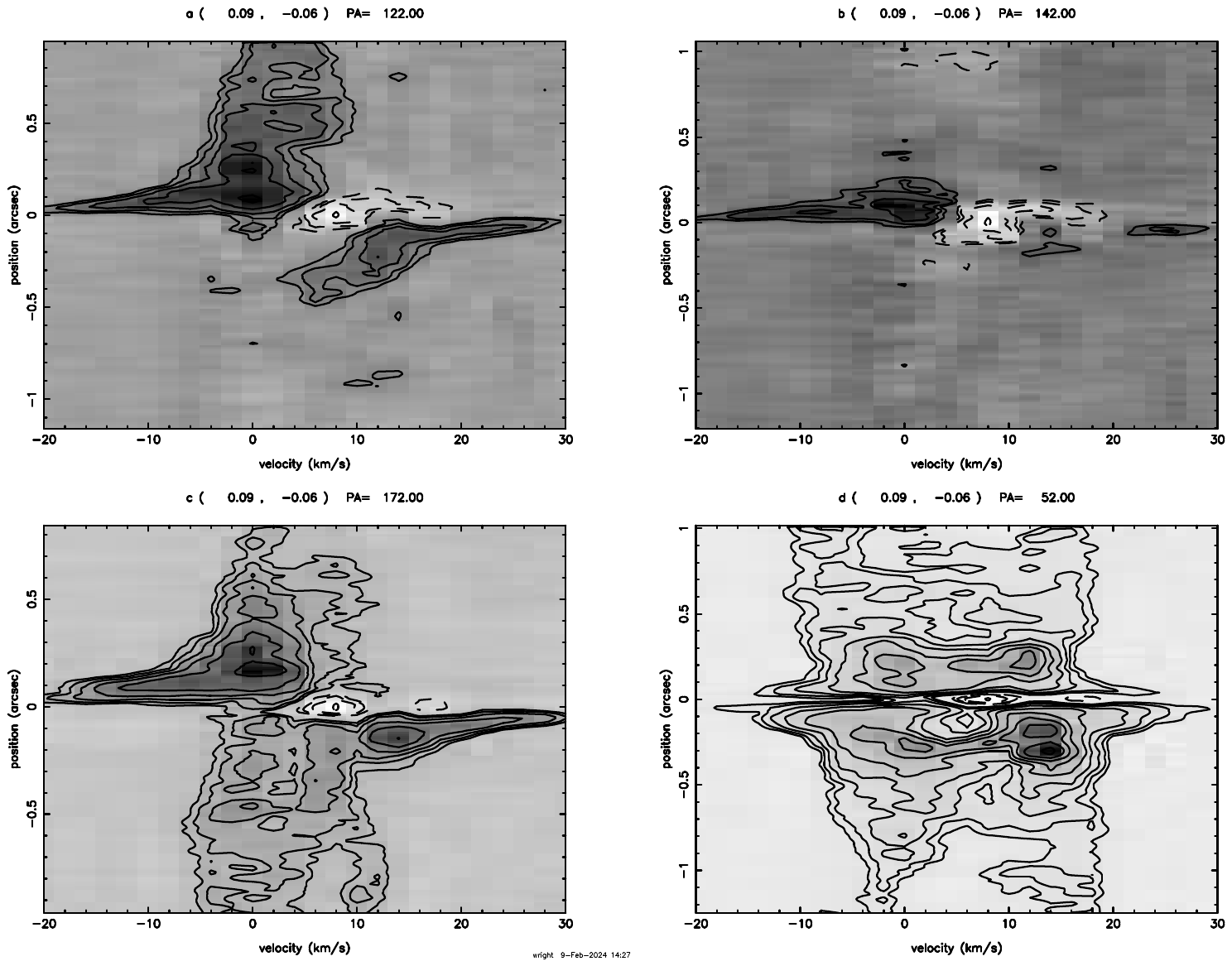}
\caption{
{\bf{Position-Velocity contours for \ce{^{29}SiO} emission.}}
Position-Velocity contours for the
214.39 GHz \ce{^{29}SiO} $v=0$ $J=5-4$ emission along the outflow edges in position angles 122 and 172\degr, along the circumbinary disk major axis in position angle 142\degr, and along the outflow axis in position angle 52\degr. Multiple velocity components are seen along the outflow edges, e.g. the emission at 10 to 12 \kms in PA 172\degr in panel c.
\label{fig:29sio.4pv}}
\end{figure*}

\autoref{fig:29sio.4pv} shows position-velocity profiles for \ce{^29SiO} emission along these 4 cuts. Two of these cuts (PA = 122$\degr$ \& 172$\degr$) are along the edges of the outflow. The line-of-sight velocity widths decrease from the full width of the \ce{^29SiO} emission close the the disk (-20 to +30 km s$^{-1}$ at 0\arcsec) to $\sim$ --5 to +15 km s$^{-1}$ at $\sim$ 0.2 to 0.3\arcsec (80 to 120 AU) from the disk.

The position-velocity profile along the disk's major axis (PA $=142\degr$) follows the rotation of the disk with strong absorption at red-shifted velocities. The position-velocity profile along the disk minor axis (PA $=52\degr$) measures the radial expansion velocity, with small contributions, $v_{z}\times cos(i)$, from the outflow velocity along the minor axis, and from the rotation velocity. The line of sight velocity along the disk's minor axis is $\sim$ --5, and $\sim$ +15 km s$^{-1}$ from the front and back side of the outflow.

\subsection{Entrainment of ambient material into the Molecular Outflow ?}

\begin{figure*}
% trim left bottom right top
\includegraphics[width=1.0\columnwidth, clip, trim=1cm 1.1cm 1.8cm 1cm]{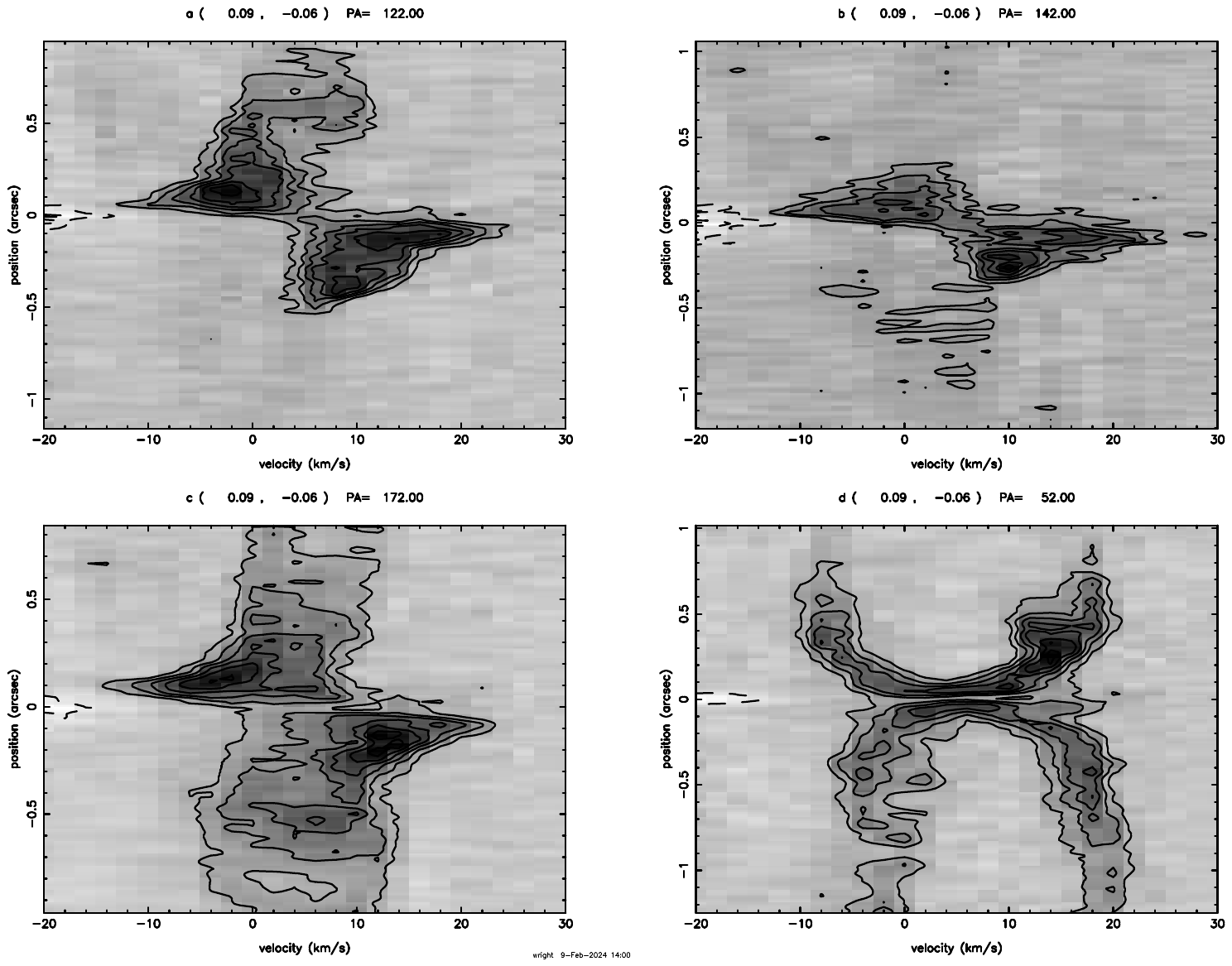}
\caption{
{\bf{Position-Velocity contours for SiS emission.}}
Position-Velocity contours for
the 217.82 GHz SiS $v=0$ $J=12-11$ emission along the outflow edges in position angles 122 and 172\degr, along the circumbinary disk major axis in position angle 142\degr, and along the outflow axis in position angle 52\degr. Multiple velocity components are seen along the outflow edges, e.g. the emission at 10 \kms in PA 122\degr in panel a.
\label{fig:sis.4pv}}
\end{figure*}

\autoref{fig:sis.4pv} shows position-velocity profiles for \ce{SiS} emission along these 4 cuts. The \ce{SiS} emission shows a more complex structure than \ce{SiO}.
Along the edges of the outflow (PA = 122$\degr$ \& 172$\degr$), the line-of-sight velocity range is consistently centered around the the systemic velocity of SrcI ($\sim$ +5 km s$^{-1}$). The velocity width decreases from its maximum ($\sim$ --14 to +24 km s$^{-1}$) close the the disk to $\sim$ --5, and +15 km s$^{-1}$ at $\sim$ 0.4\arcsec (160 AU) from the disk. 
The line of sight velocity along the edges of the outflow measures the rotation velocity, $v_{\phi}\times sin(i)$, where $i$ is the inclination of the disk to the line of sight. Assuming a disk inclination 80 to 85\degr, there are small contributions from the radial expansion velocity, $v_{r}\times cos(i)$, and the outflow velocity along the minor axis, $v_{z}\times cos(i)$. Both \ce{^29SiO}
and SiS show that the rotational velocity decreases with distance from the disk, consistent with the results shown by \citet{Hirota2017} for the Si$^{18}$O $J=12-11$ line.

 Multiple velocity components can be seen along the edges of the outflow in \autoref{fig:29sio.4pv} and \autoref{fig:sis.4pv}.

The position-velocity profile along the disk major axis (PA = 142\degr) follows the rotation of the disk. The \ce{SiS} is not seen in absorption along the major axis of the disk.

The position-velocity profile along the disk minor axis (PA = 52\degr, lower-right sub-image of \autoref{fig:sis.4pv}) measures the radial expansion velocity, with smaller contributions from the outflow velocity along the minor axis, and the rotation velocity. The line of sight velocity in \ce{SiS} along the disk minor axis increases in a parabolic fashion, i.e. more rapidly close the the disk. Gaussian fits to the spectra along the minor axis show that the line of sight velocity of the \ce{SiS} emission is initially centered at the system velocity at regions close to the disk ($\sim$ +5 km s$^{-1}$ at $<$ 0.1\arcsec (40 AU) from the disk). This diverges into two distinct, much thinner velocity structures in either direction away from the disk. The NW outflow has velocity structures centered at $\sim$ --8 and $\sim$ +17 km s$^{-1}$ (at $\sim$ +0.3\arcsec). The SE outflow has structures centered at $\sim$ --4 and $\sim$ +17 km s$^{-1}$ (at $\sim$ -0.3\arcsec). As expected, this two-pronged velocity structure is due to the contributions of the far and close sides of the outflow.

The radial velocity is $\sim$ 4 km s$^{-1}$ more blue-shifted on the NW side of the disk. Assuming the same opening angle of the outflow on both sides of the disk, this difference in the line of sight velocity along the disk minor axis measures the outflow velocity, $v_z \times cos(i)$, where $i$ is the inclination of the disk to the line of sight. Assuming a disk inclination 80 to 85\degr, the implied outflow velocity, $v_z\sim$23 to 45 km s$^{-1}$.

For comparison, \citet{Hirota2017} found, using the 484 GHz \ce{Si^18O} line, a radial velocity ($v_r$) between $\sim$ 8 to 12 km s$^{-1}$ and a rotation velocity ($v_{\phi}$) which decreases from $\sim$ 7 km s$^{-1}$ at 50 AU from the disk to $\sim$ 2 km s$^{-1}$ at 150 AU from the disk.
%}

The relative velocity of the ambient material is roughly parallel to the plane of the disk and perpendicular to the direction of the outflow. In consequence, the wide-angled nature of the outflow poses a significant disruption to the flow of the ambient material around SrcI.

Additionally, the outflow appears to be co-moving with the disk. It is not swept back by ram pressure from its motion through the medium. Material ejected from the disk earlier in its trajectory seems to move in tandem  with the disk. Naturally, this raises the question of how this ejected material is maintaining its momentum and connection to the central disk.

A possible explanation is that the outflow is being stiffened by magnetic fields \citep{Hirota2020}. It is also possible that the outflow's density, when compared to the external medium, is high enough to render the effect of motion negligible.
The outflow from SrcI has a larger cross section than the edge-on disk, and intercepts the ambient medium which it is moving through.
Assuming an ambient density, n[\ce{H2}] $\sim$1.6 $\times$ 10$^5$ cm$^{-3}$ \citep{Snell1984}, a velocity 10 \kms, ~and cross sectional dimensions of 1000 $\times$ 200 AU for the SrcI outflow, the ambient mass flow is $\sim$ 10$^{-5}$ M$_{\odot}$$\mathrm{yr^{-1}}$, or $\sim$5 $\times$ 10$^{-3}$ M$_{\odot}$ over the $\sim$ 550 years since the BN/SrcI encounter. This is comparable to the estimated mass in the outflow, which ranges from $10^{-5}-10^{-3}$ M$_\odot$ $\mathrm{yr^{-1}}$. These estimates may be affected by the motion of SrcI through the BN/SrcI explosion debris \citep{Wright2022}.

It is notable that \citet{Lopez-Vazquez2020} fitted position-velocity (PV) profiles to SiO and SiS lines and found that a model of a rotating molecular envelope accreting onto the outflow \citep{Ulrich1976} was not a good fit to the data. However, the dilution of the rotational velocity with distance from the disk (as noted above) suggests that some of the ambient material may be entrained by the rotating outflow. In laboratory experiments, \citet{Pilloch_2021} found that dry silicate dust becomes more sticky as the temperature increases, by a factor $\sim$10 up to 1000K, and then by another factor $\sim$100 for temperatures $\sim$1200K. Compression and shock heating of the envelope of the outflow may thus make adhesion and entrainment of some ambient material into the outflow more likely.

A final point of note is that the \ce{SiS} emission does not extend all the way to the disk surface, unlike the \ce{SiO} (refer to \autoref{fig:sis_mom0+1}). The parabolic increase in the radial velocity ($v_r$) at $\sim$40 - 80AU from the disk where the brightest \ce{SiS} emission is located could result from ambient material being funnelled between the disk and the outflow (refer to \autoref{fig:sis.4pv}). The velocity difference between the ambient material and the rotating outflow would result in shocks and heating of the outflow. Note that it is at this distance from the disk where the outflow expands and changes from co-rotation with the disk, and ground state \ce{SiO} masers appear in the J=1-0 and J=2-1 transitions (\cite{Greenhill2013}; \cite{Hirota2020}). 

\citet{Greenhill2013} suggested that shocks resulting from a decease in the Alfven velocity to a point below the outflow velocity may trigger the appearance of maser emission in the outflow at $\sim$100 AU. Shock heating from the flow of ambient material funnelled between the disk and the outflow provide an alternative explanation consistent with our \ce{SiS} observations. The \ce{SiS} maps the boundary of the outflow mapped in SiO, suggesting chemical interactions between the outflow and ambient material.
%}

\subsection{Chemical Interactions Between the Outflow and Ambient Material}

\citet{Li_2015} studied the sulfur chemistry in 36 sources in massive star-forming regions, observing OCS, \ce{O^13CS}, \ce{^13CS}, \ce{H2S}, and SO transitions. Their results suggest that \ce{H2S} is likely the primary sulfur-carrying molecule in massive hot cores.
The column density and abundance of OCS were generally higher than CS and SO as well. The hot core model of Orion KL could reproduce the general trend (\ce{H2S} $>$ OCS $>$ CS $>$ SO) observed in this paper. \ce{H2S} is likely to be the most abundant gas-phase sulfur-containing molecule in hot massive cores. Since hydrogenation is expected to be the most efficient process on grains, \ce{H2S} could be formed on interstellar grains \citep{van_Dishoeck_1998}. During the cold collapse phase of star formation, sulfur atoms freeze out onto grains and remain there in the form of \ce{H2S} until core heating begins. At this point, \ce{H2S} is evaporated from the grains and rapidly undergoes reactions which drive the production of SO and \ce{SO2} (\cite{Charnley_1997}; \cite{Wakelam_2005}; \cite{Wakelam_2011}). The initial destruction of \ce{H2S} by \ce{H3O+} is more efficient in high-mass protostars compared to in lower-mass ones as water is more abundant \citep{van_der_Tak_2006}. 

The observed SiS shell around the outflow may be produced in shocks from S-bearing molecular species in the ambient medium and SiO (or other Si-bearing molecules) in the outflow. Models by \cite{Pineau1993} of C-type shocks with velocities between 5 and 40 \kms and densities between $10^{4}$ and $10^{6}$ cm$^{-3}$ showed an enhancement of SO and \ce{SO_2} abundances by 2 orders of magnitude. SO decreases quickly after the passage of a shock, whilst \ce{SO_2} is enhanced during and for some time after the shock. The relative velocity of the ambient medium from the proper motion of SrcI and the rotation and expansion velocity of the disk and outflow provides a range of shock velocities between $\sim$ 0 and 30\kms.

\autoref{fig:29sio.4cuts} shows that the SiS has a complex structure, which may be evidence for this turbulent mixing. However, reactions with both atomic hydrogen and atomic oxygen will both destroy SiS rapidly. We can estimate a timescale from the SiS layer thickness and the diffusion/turbulent velocity. Estimating a thickness of 40 AU and velocity of $\sim$10 \kms from the observed SiS images, we would expect the SiS to persist for $\sim$20 yr after formation. 

Finally, we make note of a possible chemical interaction occurring between the SiO in the outflow and the SO in the ambient medium. \citet{Wright2020} found that SiS appears to map the boundary layer of the SiO outflow, and co-rotates with the SiO. This raises questions about the potential role of SO in the formation chemistry of SiS.

Several studies have examined potential formation pathways for SiS in recent years, motivated by its proposed key role in the formation of sulfide dust grains \citep{Doddipatla:2021:eabg7003}. Three pathways to forming SiS are worth consideration:

\begin{align}
    \ce{Si + SH &-> SiS + H} \label{si_sh}\\
    \ce{Si + H2S &-> SiS + H2} \label{si_h2s} \\
    \ce{SiH + S &-> SiS + H}. \label{sih_s}
\end{align}

\citet{Paiva:2020:299} have shown that Reaction~\ref{si_sh} is exothermic and barrierless. Reaction~\ref{si_h2s} is exothermic \citep{Paiva:2020:299}, and computational work by \citet{Doddipatla:2021:eabg7003} suggests that substantial SiS can be formed by this pathway as \ce{H2S} is abundant in this region. Reaction \ref{sih_s} is also exothermic and barrierless, but can produce Si + SH instead, and the branching ratios are not yet known \citep{Rosi:2018:87}. The remaining question is then the source of the reactants. 

In SrcI, with the SiS emission tracing the outer edges of the outflow, Si could be readily produced in shocks. Since the SiO outflow co-rotates with the disk close to the disk and then mushrooms out with blended rotation signature, shocks may occur as the outflow imparts angular momentum to ambient material, which creates and can be mapped in SiS.  Both SH and \ce{H2S} are similarly known to be enhanced in shocked regions, being liberated from ice mantles \citep{Neufeld:2012:L6}. 
\ce{H2S} abundances are enhanced by a factor of 1000 in the Orion hot core and the ``plateau" $[H_2S/H_2]$$ \sim10^{-6}$  relative to quiescent clouds. \citep{Minh_1990}. The large abundance of \ce{H2S} in the hot core may result from grain mantle evaporation, while the large abundance in the plateau could result from high-temperature gas-phase chemistry and/or grain-surface chemistry
\citep{Minh_1990}.

The plateau emission comes from the outflow associated with SrcI where shock chemistry may also be responsible for the large abundances of SO, $SO_2$, \ce{H2S}, and SiO. 
\ce{H2S} and $H_2^{34}S$  with $E_u<$ 500 K are seen as blueshifted absorption features at 0.2\arcsec~ resolution indicating that they originate in outflowing gas \citep{Plambeck2016}. Detailed shock models \citep{vanGelder_2021} found that SO and $SO_2$ are good tracers of accretion shocks. Desorption of SO and $SO_2$ ices can occur in high-velocity ($>$5 \kms) shocks at high densities ($>10^7$ cm$^{-3}$). They found that the abundances of atomic S and O, and in ices such as $H_2S$, $CH_4$, SO, and $SO_2$ play a key role in the abundances of SO and $SO_2$ that are reached in the shock.
Reaction~\ref{sih_s}, on the other hand, is harder to justify as important in our case. While \citet{Rosi:2018:87} show that this reaction is exothermic and barrierless, the source of sufficient SiH to produce our observed quantities of SiS is unclear. To date, SiH has only been tentatively identified in space (albeit in Orion-KL) using Caltech Submillimeter Observations \citep{Schilke:2001:281}.

Our observed coincidence of SiS with SO suggests another pathway might be viable,
\begin{equation}
    \ce{SO + SiO -> SiS + O2}. \label{so_sio}
\end{equation}

\citet{Campanha:2022:369} studied the reverse reaction,
\begin{equation}
    \ce{SiS + O2 -> SiO + SO}, \label{sis_o2}
\end{equation}
which they showed possesses a substantial activation barrier.  While this might suggest the forward reaction is efficient, in fact their potential energy surface (their Fig. 3) indicates that both forward and reverse reactions possess entrance barriers, and that the forward reaction to form SiS is substantially endothermic, but could be driven by shock heating. These endothermic reactions may prove to be important in the brief, high-temperature regime of interstellar shocks (Burkhardt et al. 2019).  Extensive modeling work would be needed to determine the extent of the effect on the overall chemistry. Further computational or experimental work exploring other potential pathways between these species is warranted.  

One potential reaction that has not been, to our knowledge, well studied is that between SiH and SH:
\begin{equation}
    \ce{SiH + SH -> SiS + H2}.
\end{equation}
There also remain some open questions about the relationship between SiS and \ce{H2O}, which may be especially relevant given the excited \ce{H2O} present at the base of the outflow in SrcI. Taken as a whole, and given the spatial overlap of many of these molecules in SrcI, further observations at high-resolution in SrcI particularly targeting SiS, SiO, SiH, SH, \ce{H2S}, SO, \ce{SO2}, and \ce{H2O} are warranted.

\subsection{\ce{H2O} in the Outflow}
%{\color{magenta}
The origin of this water line in the gas phase is not clear. The \ce{H2O} emission extends deeper and likely formed in the disk, but icy grain mantles are unlikely to have survived in the hot disk for $\sim$550 years since the BN/SrcI encounter. Hydrated minerals are another possible source of \ce{H2O}. The AlO line emission that was mapped in the outflow close to the disk suggests that refractory grain cores and grain mantles are destroyed \citep{Wright2020}. Carbon grains are destroyed at $\sim$800 - 1150 K, silicate grains are evaporated at $\sim$1300 K, and AlO at $\sim$1700 K \citep{Lenzuni1995}.

This vibrationally excited water line,
% H2O 55, 0-64, 3 ν2 = 1 (Eu = 3461.9 K) 
is also detected in the Keplerian disc around the O-type protostar G17.64+0.16 \citep{Maud_2019}
and in the O-type protobinary system IRAS 16547–4247 \citep{Tanaka2020}. The \ce{H2O} $v_{2}$ = 1 emission with E$_{u}$ = 3464 K is concentrated at the positions of protostars \citep{Tanaka2020} 
and in several other high mass protostellar objects \citep{Ginsburg2023ApJ}.

Near IR observations of \ce{H2O} toward the massive protostar AFGL 2136 IRS 1 identified 47 ro-vibrational transitions 
in warm (500 K),  dense (n(H${_2}$) $>$5 $\times 10^{9} cm^{-3}$) gas, suggesting an origin close to the central protostar \citep{Indriolo_2013}.

\begin{figure}
% trim left bottom right top
\includegraphics[width=1.0\columnwidth, clip, trim=2.5cm 3.7cm 2.5cm 1cm]{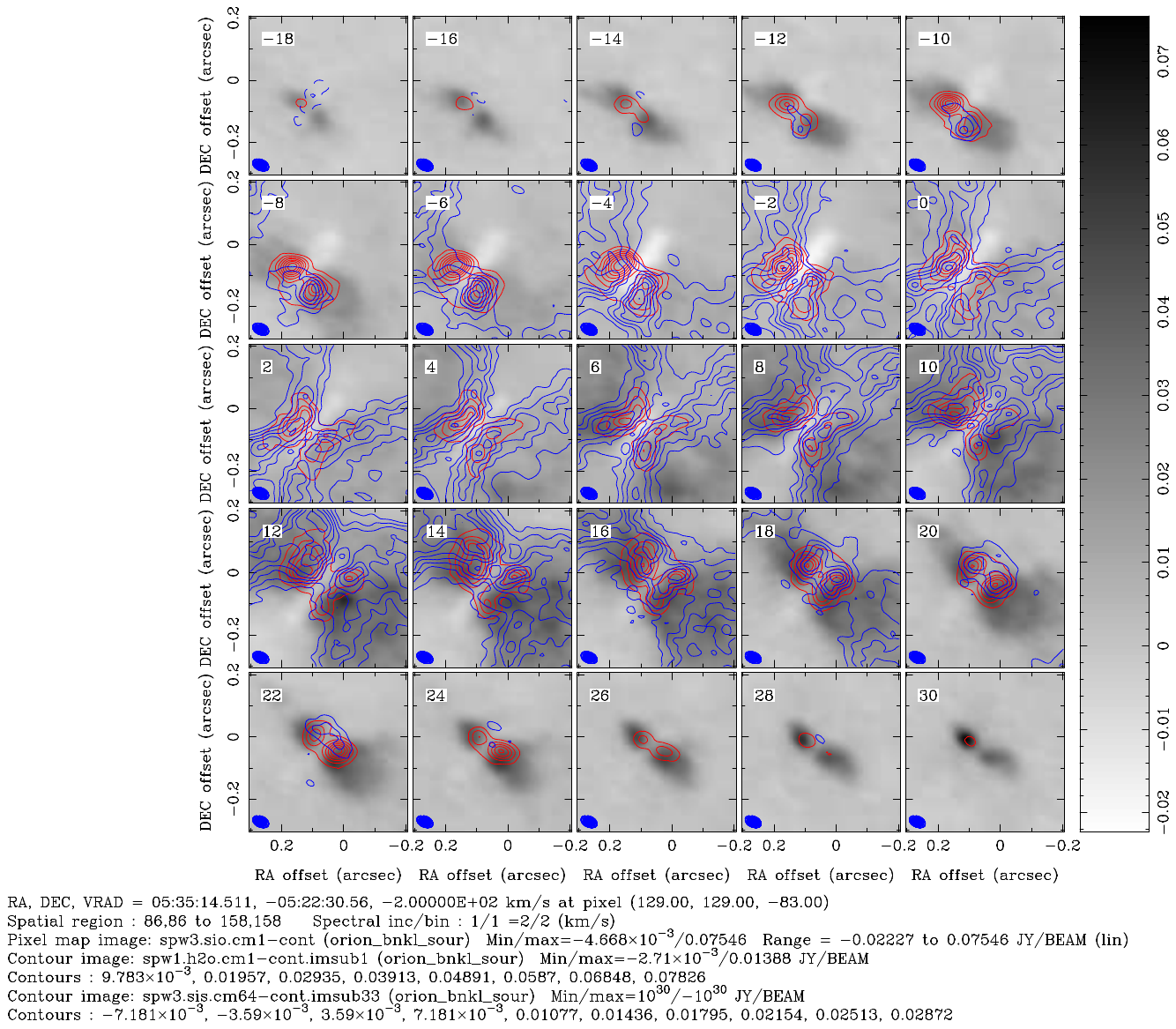}
\caption{
{\bf{Structure and kinematics of \ce{H2O}, SiS, and SiO emission close to SrcI.}}
This figure shows the 232.69 GHz \ce{H2O} emission (red contours), 217.82 GHz SiS emission (blue contours), and 217.10 GHz SiO $v=0$ $J=5-4$ (pixel image) in 2 km/s channels. The SiO emission is
shown as the grey scale in units of Jy/beam as indicated in the wedge. The continuum emission from the disk has been subtracted from the images. The SiO emission is seen in absorption against the disk and is visible in white. 
\label{fig:h2o+sis+sio}}
\end{figure}

On the other hand, \autoref{fig:h2o+sis+sio} shows that  \ce{H2O} emission peaks are associated with SiS close to the disk, suggesting that heating by shocks may play a role in the formation and dissociation of \ce{H2O}, $\sim$50 AU from the SrcI disk, where the rotating SiO column expands abruptly into a turbulent wide angle outflow. These distributions fit into a scenario where the SiO is formed from grain destruction following the dissociation of \ce{H2O} \citep{Schilke1997}. In these models, grain mantles and grain cores are destroyed in shocks; Si is released into the gas phase, and then oxidized by dissociation products of \ce{H2O}.

\citet{Ginsburg2018} fitted a Keplerian rotation curve to the envelope of a position-velocity along the major axis of the SrcI disk. However, the butterfly-like shape seen in \autoref{fig:nacl_mom0+1} suggests that the \ce{H2O} also has an outflow component.

\autoref{fig:h2o+sis+sio} compares the \ce{H2O} emission with the SiS and SiO at $\sim50\times30$ mas resolution. At velocities -18 to -10 and 22 to 30 \kms, the \ce{H2O} traces the SiO outflow along the minor axis of the disk, while at velocities -8 to 20 \kms, the \ce{H2O} traces the bright \ce{SiS} emission in the inner part of the SiS outflow.

\subsection{Possibility of Bondi-Hoyle-Lyttleton Accretion onto the Circumbinary Disk of Source I}

As summarized by \citet{Edgar_2004}, the Bondi-Hoyle-Lyttleton  (BHL) accretion model describes a possible process of for accretion onto a moving point mass passing through a uniform gas cloud. In essence, the gravity of the mass lenses the medium into a trail in its wake. From this dense, aligned column, the mass can accrete. This model neglects pressure and thermal effects, such as the trapping of thermal energy in the trailing column. Accretion onto the disk from the ambient cloud would be limited to roughly the BHL rate, which for $n_{H_{2}} = 1.6 \times 10^5 \; \text{cm}^{-3}$, $\rho \sim 5 \times 10^{-19} \; \text{g} \; \text{cm}^{-3}$, $M = 15 \; M_\odot$, and $v= 10 \; \text{km} \; \text{s}^{-1}$.
relative to the interstellar background is given by

\begin{equation}
    \dot{M} = \lambda_* 4 \pi \rho \frac{\left(GM\right)^2}{\left(v^2 + c_s^2\right)^{3/2}} \approx 5 \times 10^{-7} M_\odot \; \text{yr}^{-1}
\end{equation}

\noindent where $c_s = \sqrt{ k_B T / \mu m_p }$ is the ambient thermal (sound) speed $\sim$ $1\text{-}5 \; \text{km} \; \text{s}^{-1}$  for $T=10\text{-}1000 \; \text{K}$, and $\mu = 2.8 $. $\lambda_*$ is a dimensionless constant of order unity \cite{shu1991physics}.

The SO line is a possible tracer of this trailing shocked ambient medium around SrcI resulting from the BN/SrcI explosion. \citet{Bally2017} observed the 219.95 GHz SO line with ALMA at 1\arcsec ~resolution. Figure 17 from \citet{Wright2020} shows a 30\arcsec ~region image centered on SrcI. The SO is seen throughout this region with a similar filamentary structure in the SiO and CO emission.

\autoref{fig:son344GHz} shows SO emission around SrcI, and the expected proper motion of SrcI in 100 years. Table 1 from \citet{Rodgriguez2017} gives the positions and proper motions of the runaway stars in OMC1. The proper motion vector of SrcI has a PA $=$152 $\pm$ 4\degr. It appears to be misaligned by  9$\pm$4\degr with the SrcI disk major axis (PA$=-$37 $\pm$ 0.4\degr, as observed at 220-340 GHz)  \citep{Wright2020}. The SrcI major axis aligns with the BN proper motion (323 $\pm$ 2{\bf{\degr}}) \citep{Rodgriguez2017}, within 2{\bf{\degr}} .
This is expected, as in the BN-SrcI interaction, material in orbit around the stars experienced a torque that align their disks
with the stellar proper motion vectors \citep{Bally2020}. The small misalignment of the SrcI disk is consistent with the hypothesis that
the BN-I interaction shaped the disk rather than subsequent BHL accretion. Misalignment of the SrcI disk with its proper motion may produce the turbulent wake seen in SO emission. It is also possible that this misalignment is the product of source x and the current SrcI binary, or merger remnant, having separated after the initial SrcI disk formation and relaxation.

Since the outer radius of SrcI's disk is smaller that the gravitational radius, $r_G$ $\sim GM/v_{rot}^2$ $\sim$ 130 AU, it is likely that most of the SrcI disk has had to re-form since the dynamic interaction $\sim$550 years ago from pre-existing circumstellar material. This formation process may be contributing to the current accretion (and outflow). The fall-back and resulting turbulence might contribute to shocks, and enhanced accretion from disk to star.

\begin{figure}
% trim left bottom right top
\includegraphics[width=1.0\columnwidth, clip, trim=3cm 4.0cm 2.5cm 1cm]{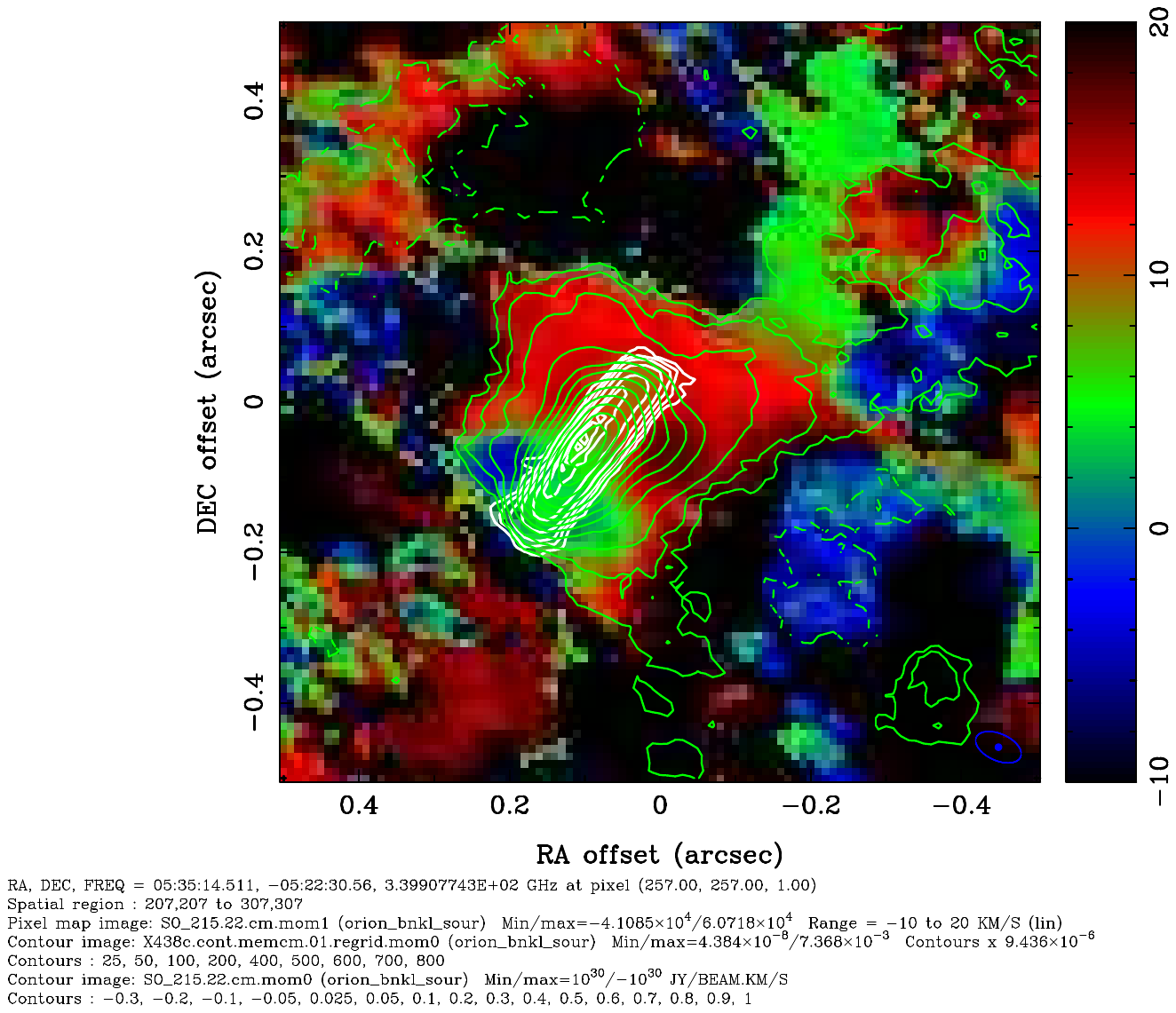}
\caption{
{\bf{ SO 215.22 GHz emission  with uniform weighting of the u-v data to emphasize the fine scale structure.}}
The color image shows the moment 1 velocity image with velocities from -10 to +20 \kms indicated in the color wedge. 
Green contours show the SO emission, integrated over velocity range -20 to +30 km/s. Contour levels are at -0.3, -0.2, -0.1, -0.05, 0.025, 0.05, 0.1, 0.2, 0.3, 0.4, 0.5, 0.6, 0.7, 0.8, 0.9, 1 Jy/beam. The peak brightness with 1 km/s velocity resolution is 744 K. The synthesized beam FWHM (64$\times$36 mas, PA 66\degr) is indicated in blue in the lower right.
The white contour map shows 340 GHz continuum emission at levels 25, 50, 100, 200, 400, 500, 600, 700, 800 K. The continuum image has been convolved by a 10 mas FWHM beam.
%Maximum Entropy image.
\label{fig:so}}
\end{figure}

\autoref{fig:so} shows the $215.221$ GHz SO emission
with uniform weighting of the u-v data to emphasize the small scale structure with a velocity resolution of $1$ \kms.

The SO line shows an envelope of emission around the disk, with a steep gradient at the SE end of the disk, and an extended tail to the NW of the disk. There are also significant emission features at 0.2 to 0.4\arcsec to the NE and SW of the disk.
The emission surrounding the disk is largely red-shifted, consistent with the 0.2\arcsec resolution spectra of SrcI plotted by \citet{Plambeck2016}, in which the sulfur-bearing species (CS, \ce{H2S}, SO, and \ce{SO2}) exhibit prominent blue-shifted absorption profiles. 

SO and \ce{SO2} are heavily absorbed across SrcI at velocities of -8 to +4~\kms \citep{Wright2020}. This blue-shifted absorption is the reason for the domination of red-shifted emission around the disk.
Since the velocity width of the absorption is comparable with the halfwidth of the SiO emission lines, it is likely that the absorbing molecules are located in the (cooler) outer layers of the SrcI outflow, rather than in unrelated foreground gas \citep{Plambeck2016}.

In \autoref{fig:so}, the $215.221$ GHz SO$\quad v=0\quad(J=5-4)$  line shows a steep gradient at the SE end of the disk, with an extended tail to the NW of the disk.
The SO suggests a bow shock from the motion of SrcI through the medium, with a turbulent wake similar to the simulations in \citet{Moeckel_2009}.
\ce{SO} is heavily present in the region around the disk, and very roughly follows a velocity pattern around the disk. The leading (SE) side of the disk has blue-shifted SO emission, while the leeward side has red-shifted SO emission. 
This would be expected if SO was wrapping around the disk between the outflow lobes; about half the incoming SO would
have a velocity component towards us along our line of sight to the disk, and red-shifted towards the leeward side of SrcI.

\begin{figure}
% trim left bottom right top
\includegraphics[width=1.0\columnwidth, clip, trim=2.5cm 3.6cm 2.7cm 1cm]{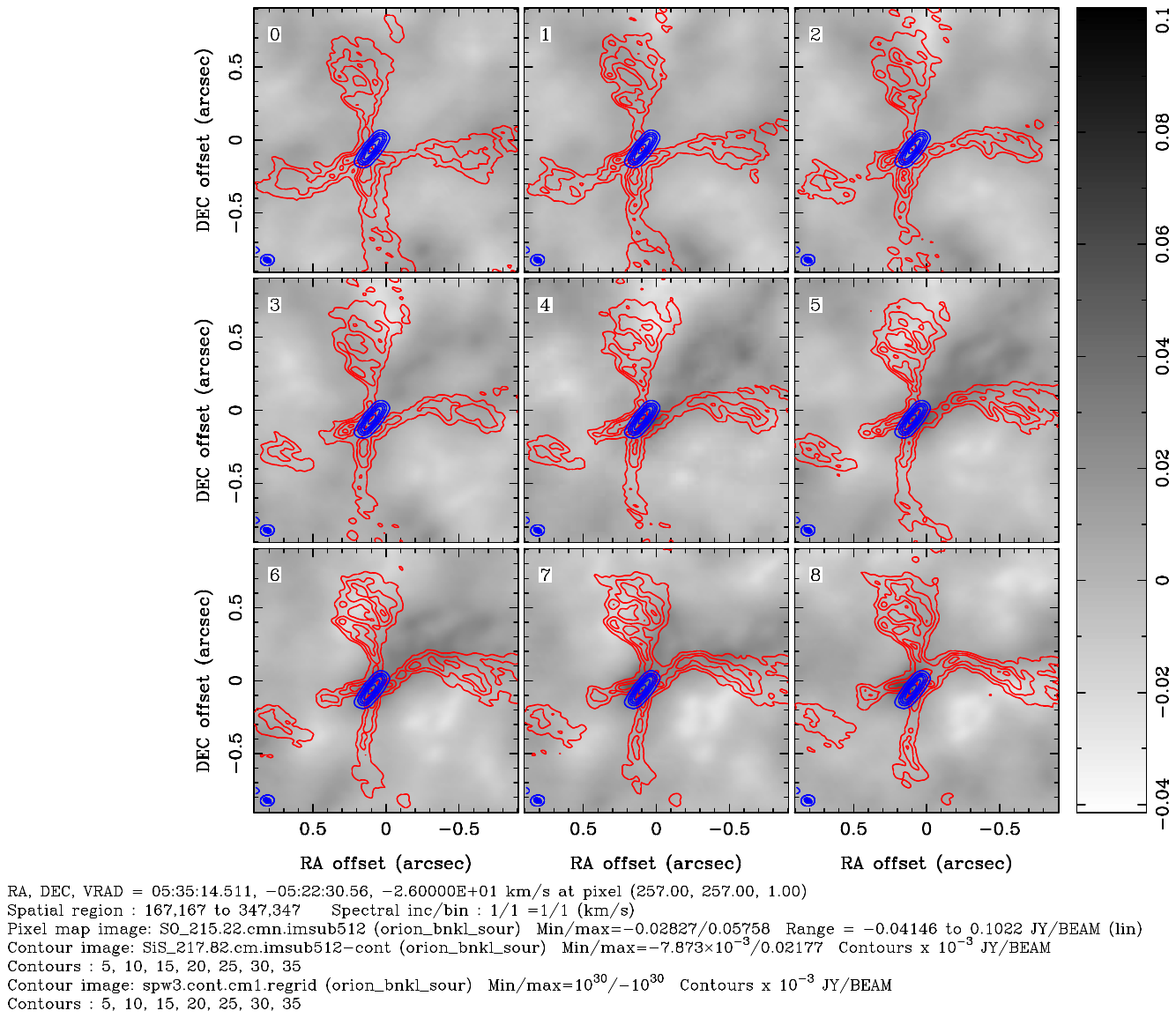}
\caption{
{\bf{ SO and SiS emission in 1 km/s channels}}. The grey pixel image shows the 215.22 GHz SO emission with channels in 1 km/s intervals.  The uv-data was naturally weighted to enhance the brightness sensitivity. The wedge shows the SO brightness in units of Jy/beam. The RMS noise = 9 mJy/beam (35 K). The synthesized beam FWHM (95$\times$76 mas, PA -83\degr) is indicated by the blue outline in the lower left.
The red contours map the 217.82 GHz SiS emission. Contour levels are at 5, 10, 15, 20, 25, 30, and 35 mJy/beam. The blue contours map the 218 GHz continuum emission from the SrcI disk. Contour levels are at 5, 10, 15, 20, 25, 30, and 35 mJy/beam. The synthesized beam FWHM (54$\times$34 mas, PA 65\degr) is indicated in solid blue in the lower left.
\label{fig:SiS+SO.cmn}}
\end{figure}

Past this region, we see a significant body of SO extending away from the disk that has a small $v_{los}$ component (indicated in green, where 0 $<$ $v_{los}$ $<$ 10 \kms). 
\autoref{fig:SiS+SO.cmn} shows the velocity structure of the trailing plume of 215.22 GHz SO emission to the NW of the disk, using  natural weighting of the uv-data to enhance the brightness sensitivity in 1 \kms intervals. The RMS noise level is 9 mJy/beam (35 K) in a synthesized beam FWHM 95$\times$76 mas at PA -83\degr.
 
This region may correspond to the ``quasi-static" body mentioned above. Material within a radius 
$R_A = 2GM/\sqrt(c_s^2+ v^2)$ $\sim$90 au for $M\sim$ 15 $\msun$, will be accreted on a time scale
$R_A/c_s$ $\sim$ 300 yr \citep{Moeckel_2009}, where $c_s = \sqrt{ k_B T / \mu m_p }$ is the ambient thermal (sound) speed $\sim$ 1.4 \kms for T$\sim$500 K.

In addition, studies have shown using simulations that the radiation pressure exerted by massive protostars does not preclude accretion; mass can be delivered into the star through non-axisymmetric disks and filaments that self-shield against radiation, created by gravitational and Rayleigh-Taylor instabilities while radiation escapes through optically thin bubbles \citep{Krumholz2009}. Gravitational instabilities can also lead to the formation of companion stellar objects \citep{Krumholz2009}.

Another difficulty of getting material to infall upon massive stars is removing the angular momentum from the accreting material. The accretion disk is a mediating structure for this process, providing a buffer between the surrounding material and the protostar. For further discussion of this topic, see \citet{Pudritz_2019}. 

The material accretes onto the disk as it loses some angular momentum, and then more angular momentum is removed through processes such as D-winds and X-winds \citep{Shu2000}, resulting in material falling onto the protostar from the disk. These steps are time-variable and may not always occur at the same rate relative to one another.
X-winds are an alternative mechanism for generating outflows from disks. According to this model, high velocity outflows come from a narrow annulus where the star magnetosphere interacts directly with the inner edge of the disk, at the co-rotation radius \citep{Shu2000}.
The hot spot observed on the inner edge of the SrcI disk \citep{Ginsburg2018} may result from this mechanism, or alternatively from the proximity of a binary protostar. If SrcI is a binary protostar, it is not clear how the magnetospheres of the two stars interact with each other, and with the disk.

Observations of the scattered near-infrared emission suggest that the near-infrared reflection spectrum observed in the Orion-KL region is produced close to SrcI, and similar to stellar photospheres in the range Teff $\sim$ 3500 to 4500 K \citep{Testi2010}.
The spectrum excludes SrcI being a massive protostar rotating at breakup speed. \citet{Testi2010} suggest that the absorption spectrum comes a disk surrounding a $\sim$10~$M_{\odot}$
protostar, accreting from its disk at a high rate of a few $\times 10^{-3}$ $M_\odot$/yr.

\citet{Hosokawa2010} studied the  evolution of massive protostars with disk accretion. They identified four evolutionary phases, including deuterium burning which may account for a significant fraction of the luminosity of SrcI at some point. However their models show that that a massive protostar bloats to a radius of several hundreds of $R_{\odot}$ with an accretion rate exceeding 10$^{-3}$ $M_\odot$/yr. The large radius leads to a low effective temperature consistent with that observed for SrcI. They also note that the low UV luminosity will inhibit the formation of an H II region until mass accretion onto the star deceases significantly. 
Assuming a stellar radius of 1 AU with color temperature 4000K, the theoretical luminosity would be, 

$$L_{tot} = 4 \pi R^{2} \sigma T^{4} \simeq 1.01 \times 10^{4} L_\odot$$
%> >>> 5.6e-5*4*3.14*(1.5e13)**2*4000**4/4e33

Comparatively, if we assume an infall rate $5\times10^{-3}$ M$_\odot$ / year onto a central mass of 15 M$_\odot$, the luminosity to be

$$L_{tot} = \frac{G M \dot{M}}{R} \simeq 1.05 \times 10^{4}L_\odot $$

Currently, no direct observations of infall onto SrcI have been made. The expected infall rate ($\dot{M}$) is inferred from the observed outflow to range from 10$^{-5}$ to 10$^{-3}$ $M_\odot$/yr. 
The \ce{H2O} rotation-fitting done by \citet{Ginsburg2018} favor a $\sim$15~$M_{\odot}$ protostar with a high accretion rate.
To maintain the observed luminosity of $\sim$10$^4$ $L_\odot$, if the accretion onto the circumbinary disk is akin to the Bondi-Hoyle rate ($\sim$10$^{-7}$ $M_\odot$/yr) and accretion from the disk onto the star is $\sim$10$^{-3}$ $M_\odot$/yr, then the disk will be depleted on a timescale of $M_{disk}$/$\dot{M}$. \citet{Plambeck2016} estimated M$_{disk}$ $\sim$ 0.02 to 0.2$M_\odot$, giving a disk depletion time of 200 to 2000 yr for $\dot{M}$$\sim$10$^{-3}$ $M_\odot/yr$.

If a higher rate of accretion onto the circum-binary disk is not occurring, then the protostar will likely deplete the disk and reach the final stage of its pre-main sequence evolution. This is a rather exciting possibility, as it would expect SrcI to become a directly visible stellar object within several centuries. 
However, the source would still be in the Orion A cloud for a while. To emerge into the Orion Nebula would take tens of thousands of years. If it is currently $\sim$ 0.1 pc behind the ionization front with a radial velocity (v$_r) \sim 4$ \kms, this would take 
$\sim 2.5~10^{4}$ years. Also, if $\dot{M} \sim 10^{-3}$ $M_\odot/yr$, the photosphere will remain on the AU scale, and after the disk is gone, it will take a Kelvin-Helmholtz time-scale to heat up from 4,000 K to $\sim$ 40,000 K for the newborn star to begin ionizing the surroundings.

\subsection{Evidence of a Warped Disk in Salt Emission}
Salt emission has been detected in nine high-mass YSO disks so far \cite{Ginsburg2019,Maud_2019,Tanaka2020,Ginsburg2023ApJ}. 
In SrcI, salt emission is found close to the dust layer at the surface of SrcI and maps the rotation close to the disk surface \citep{Ginsburg2018,Ginsburg2019}.

\begin{figure}
% trim left bottom right top
\includegraphics[width=1.0\columnwidth, clip, trim=2.5cm 3.7cm 2.5cm 1cm]{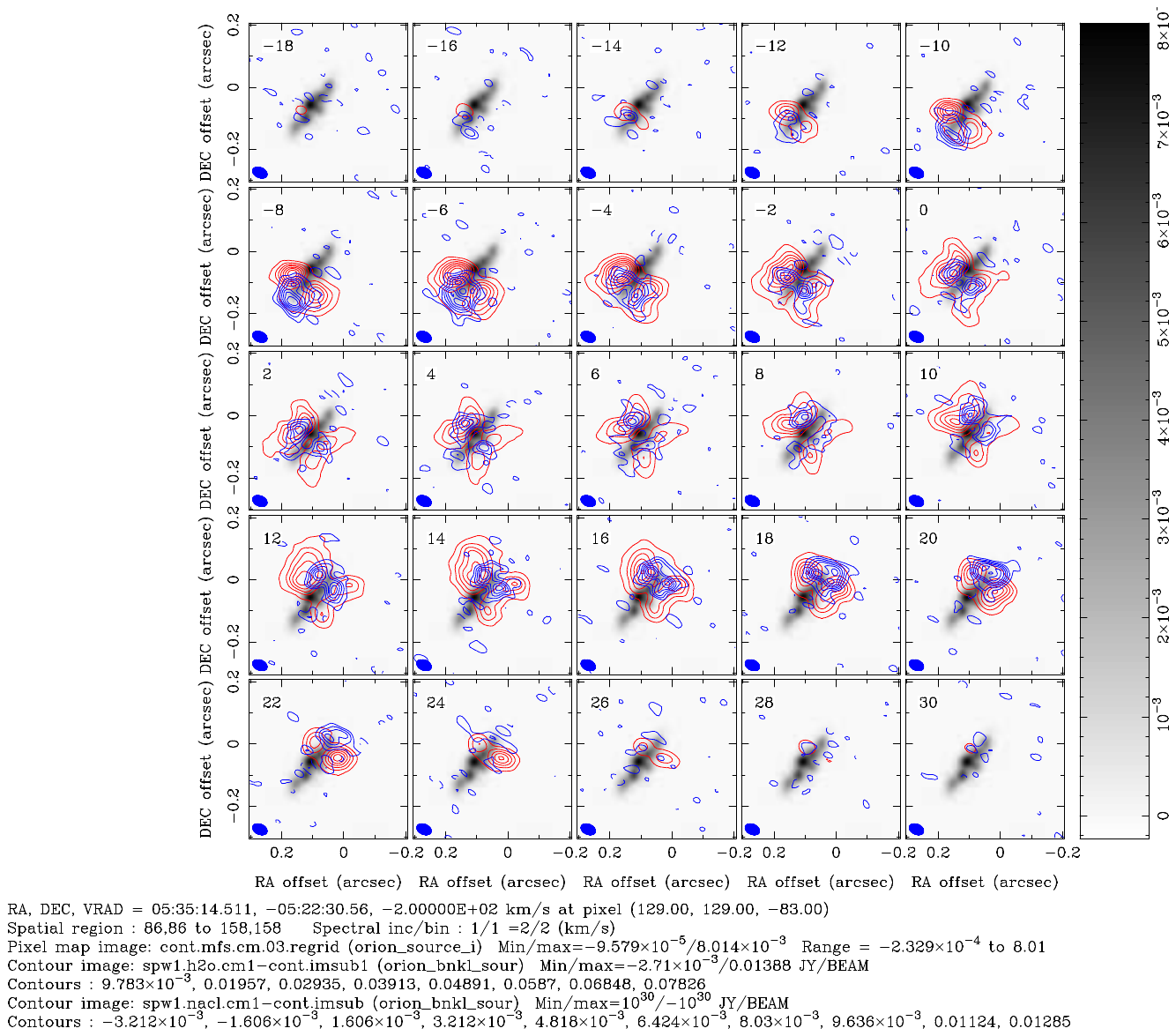}
\caption{ {\bf{\ce{H2O} and NaCl emission in 2 km/s channels.}}
Comparison of the 232.69 GHz \ce{H2O} emission
(red contours) and the 232.51 GHz NaCl emission (blue contours). 
Both NaCl, and  \ce{H2O} map the rotation of the disk.
The NaCl emission is more closely associated with the continuum disk and extends to greater radii along the disk major axis than the \ce{H2O} emission.
The synthesized beam FWHM for the \ce{H2O} and NaCl (54$\times$34 mas, PA 66\degr) is indicated in blue in the lower left.
The 99GHz continuum emission is shown as the grey pixel image
with brightness from 0 to 9 mJy/beam indicated in the wedge. The continuum data has been convolved by a 30 mas FWHM beam.
\label{fig:h2o+nacl}}
\end{figure}

A comparison of the \ce{H2O} and NaCl at $\sim50\times30$ mas resolution is shown in \autoref{fig:h2o+nacl}. The NaCl emission is more closely associated with the continuum disk and extends to greater radii along the disk major axis than the \ce{H2O} emission.
\autoref{fig:nacl_mom0+1} shows the moment 1 image--the velocity weighted intensity--of the \ce{NaCl} emission. This shows a simple, linear velocity gradient (implied by the linear color gradient across the disk), which is consistent with an almost edge on thick disk.
\citet{Matthews2010} found that the SiO masers suggest an inclination of $\sim$85$\degr$. Additionally, \citet{Ginsburg2018} fitted a Keplerian rotation curve to the envelope of a position-velocity along the major axis of the SrcI disk in good agreement. 

\begin{figure}
% trim left bottom right top
\includegraphics[width=1.0\columnwidth, clip, trim=2.5cm 3.7cm 2.5cm 1cm]{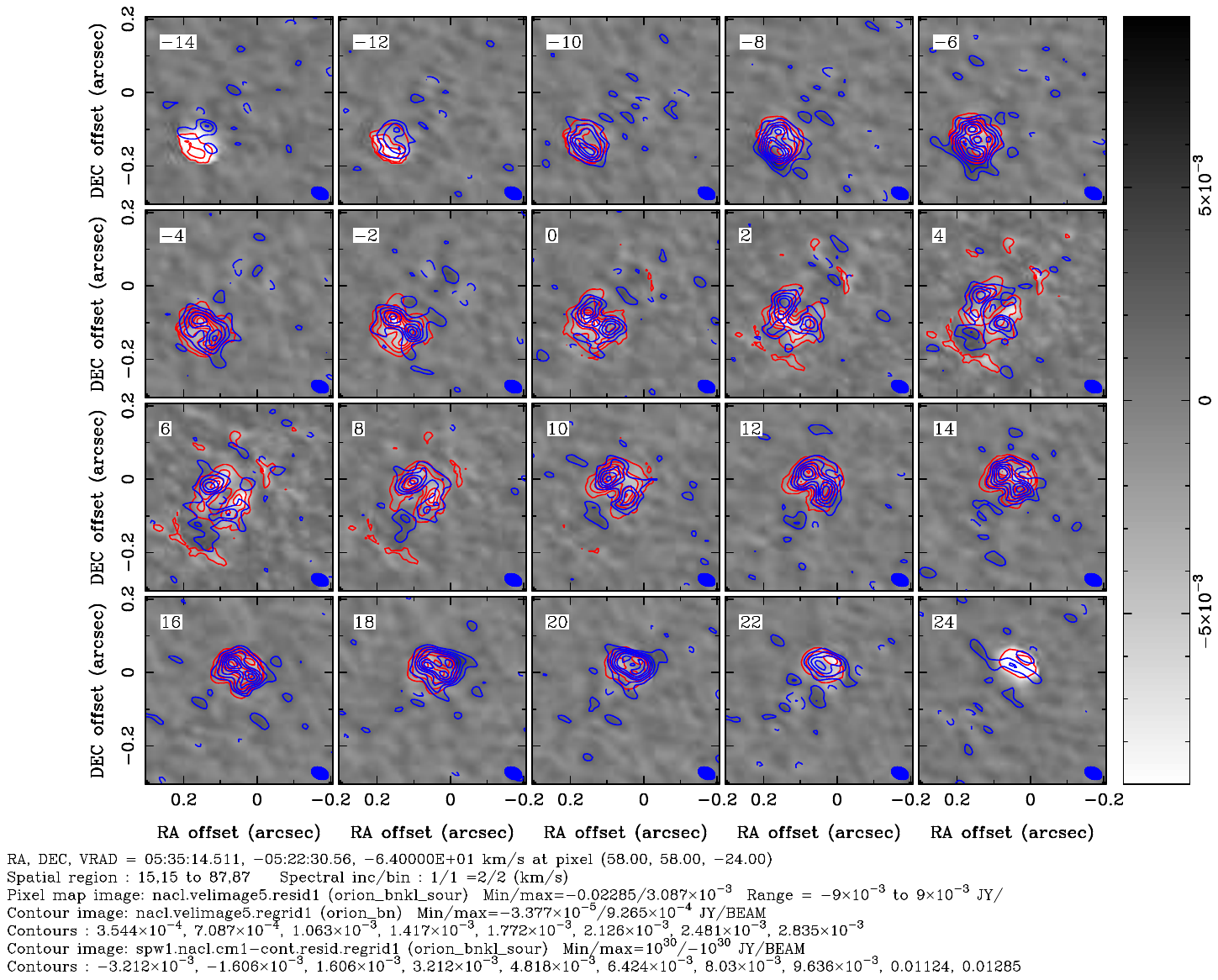}
\caption{
{\bf{ Structure and kinematics of NaCl emission.}}
Comparison of the structure and kinematics of the observed NaCl emission and a model generated from a linear velocity gradient along the disk major axis.
The blue contours show the 232.51 GHz NaCl emission in 2 km/s channels. Contour levels are at -32, -16, 16, 32, 48, 64, 80, 96, 112, and 128 mJ/beam.
The red contours show the simulated emission from a model rotation curve. 
The grey pixel image shows the difference between the model and the data in units of Jy/beam indicated in the wedge. Note the significant difference between the data and the model at large absolute velocities. The RMS error of this image is 1.2 mJy/beam. The synthesized beam FWHM (47$\times$31 mas, PA 65$\degr$) is indicated in blue in the lower right.
\label{fig:nacl.cm1-cont+velimage5.resid}}
\end{figure}

Using a linear velocity gradient combined with the observed NaCl moment 0 image, we generated model images in 2 km/s channels for comparison with the data shown in \autoref{fig:nacl.cm1-cont+velimage5.resid}. 

This simple model agrees with the observed NaCl emission in 2 km/s channels between -10 and +20 km/s. The maximum differences between the model and the data occur at the ends of the disk where the model and the data are asymmetrically offset. We used the MIRIAD tool {\bf imdiff} to fit the offset as as a function of velocity. The offsets can be minimized by allowing the position angle of the disk to change with radius.
A possible explanation for this disagreement at the velocity and radial extrema is that the NaCl emission resulted from (recently) accreted material having a different angular momentum compared to the original disk which resulted from the BN/SrcI encounter. Accreted material, or the localized outflow resulting from accreted material, could exert a torque. This may produce the warping observed in the 99 GHz continuum emission from the disk \citep{Wright2022} and the NaCl emission at the ends of the disk.
Alternatively, misalignment between the rotation axis and the B-field direction could also create a warped disk structure \citep{Machida_2019}.

\begin{figure}
% trim left bottom right top
\includegraphics[width=1.0\columnwidth, clip, trim=1.cm 3.7cm 2.5cm 1cm]{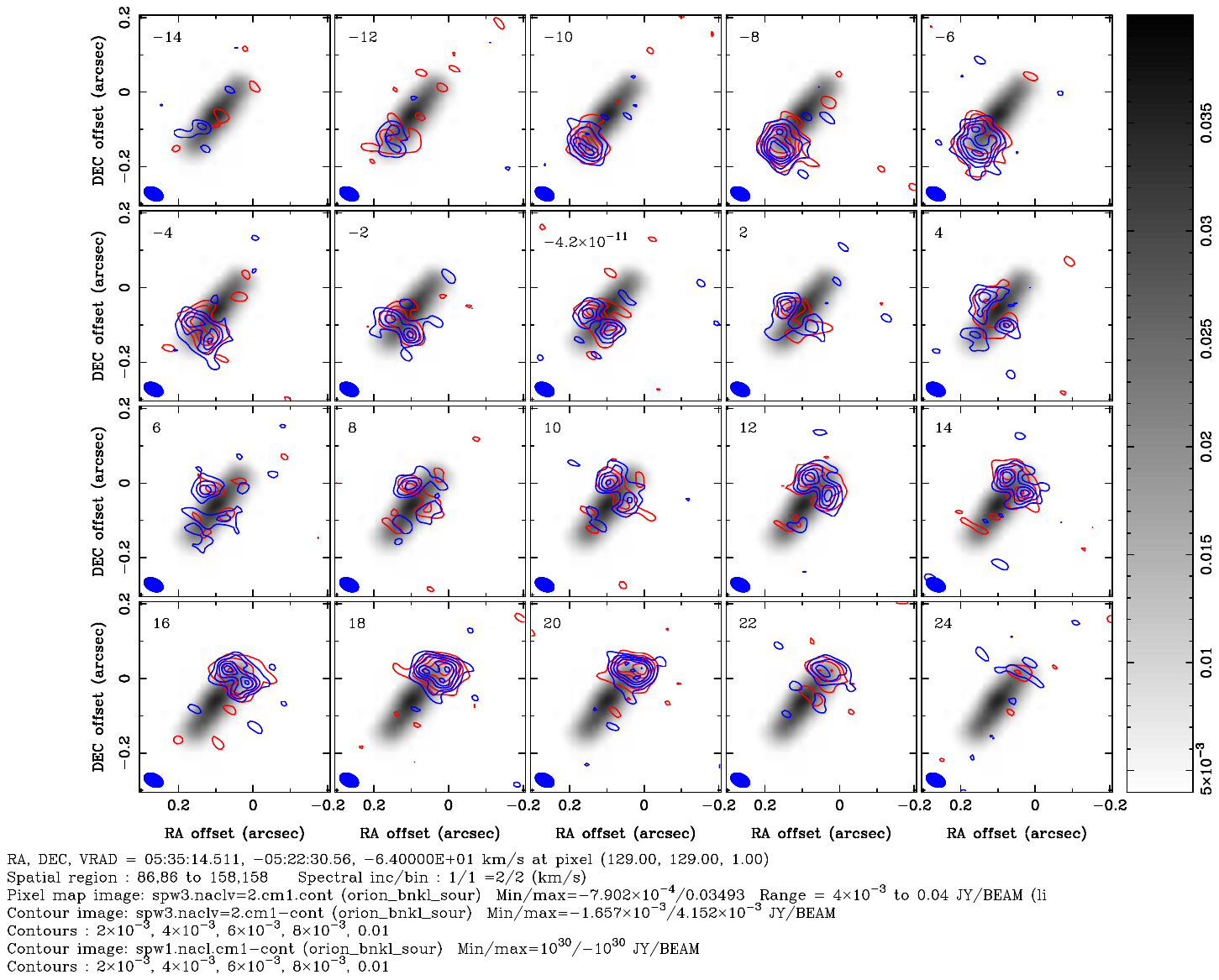}
\caption{
{\bf{Comparison of two NaCl emission lines.}}
A comparison of the 232.51 GHz NaCl $v=1$ $J=18-17$ (blue contours) and the 217.98 GHz NaCl $v=2$ $J=17-16$ (red contours) in 2 km/s channels. Note that the v=1 and v=2 emission comes from the same regions in the disk at every velocity. Contours at 2, 4, 6, 8, and 10 mJy/beam. The pixel image shows the continuum emission from the disk at 230 GHz in units of Jy/beam indicated in the wedge.
The synthesized beam FWHM is indicated in blue in the lower left.
\label{fig:naclv=1+v=2}}
\end{figure}

Additionally, the salt lines could be excited by infall of material onto the disk. \autoref{fig:naclv=1+v=2} shows a comparison of the 232.51 GHz NaCl $\quad v=1\quad(J=18-17)$ and the 217.98 GHz NaCl$\quad v=2\quad(J=17-16)$ in 2 \kms channels. Note that the $v=1$ and $v=2$ emission come from the same regions, and have the same velocity distribution, suggesting a common origin for different transitions. This common origin may be excitation due to shocks created by accretion onto the disk.

\begin{figure}
% trim left bottom right top
\includegraphics[width=1.0\columnwidth, clip, trim=3cm 4.3cm 2.5cm 1cm]{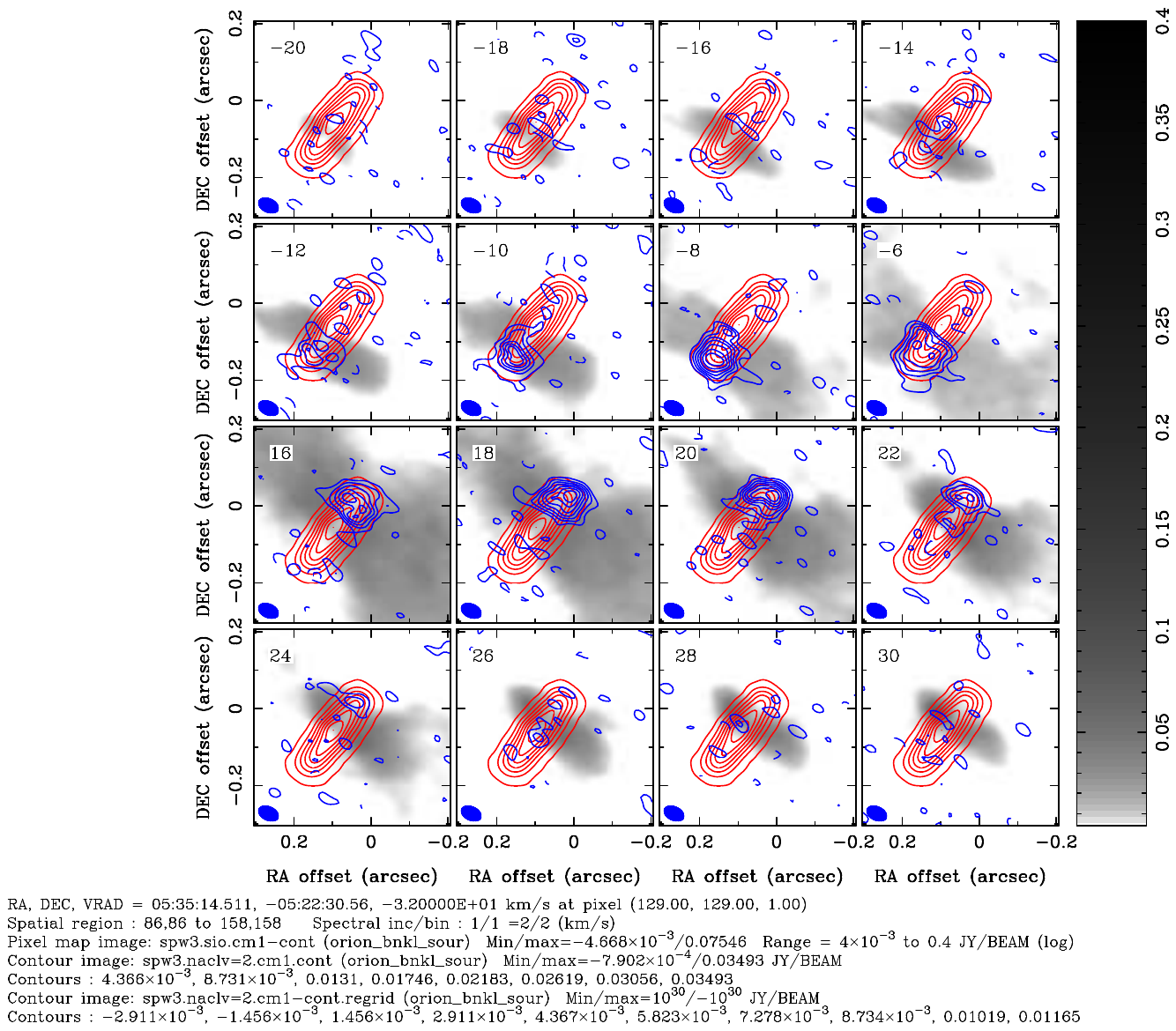}
\caption{
{\bf{ NaCl and SiO outflow.}}
Comparison of NaCl v=2 17-16 at 217.98 GHz (blue contours at intervals 1.46 mJ/beam)
%SiS (green contours at intervals 3.25 mJy/beam),
and SiO  in 2 \kms channels shown in the grey scale image in units Jy/beam on a log scale from 0.04 to 0.4 Jy/beam as indicated in the wedge. Note the NaCl emission at ends of the disk, while the SiO has footprints at a smaller radius in the disk.  The 217 GHz continuum emission is shown in red contours at intervals 4.36 mJy/beam. The synthesized beam FWHM is indicated in blue in the lower left.
\label{fig:naclv=2+sio}}
\end{figure}

\autoref{fig:naclv=2+sio} shows a comparison of 217.98 GHz NaCl$\quad v=2\quad(J=17-16)$ and SiO in 2 \kms channels. Note that the NaCl emission that comes from the ends of the disk may be a byproduct of an accretion shock of material falling into the disk, while the SiO emerges from a smaller radius in the disk, corresponding to the footprint of a magneto-centrifugal outflow from the disk. Some of the material falling towards the disk at radii within this footprint would be swept out by the outflow.
 
The symmetry of the NaCl bright spots about the rotation axis may be from the rotation of the disk. If this explanation is correct, then this implies that accretion is followed by orbital redistribution i.e. the salt must stay in the gas phase for the orbital timescale at $\sim$ 30 AU  radius for a 15 $M_\odot$ star, $>$40 years.

The vibrationally excited NaCl must be populated on these time scales because of the high Einstein A-values \citep{Ginsburg2019}. However, the density and temperature rise within 100 - 1000 yr as the shock goes through the surface layers of the disk \citep{Burkhardt_2019}. The 1/e scale height of the disk surface estimated from the 340 GHz continuum is a only a few au \citep{Wright2020}, and presents a wide range of temperatures and densities for accreting material. More detailed modeling is needed to see if accretion shocks could populate the observed NaCl levels.

Similar to the NaCl distribution, observations of the  $5_0 - 4_0$ (A) methanol line at 241.79143 GHz, ($E_u$ 35K) in the surface layer at the ends of the disk in the Class 0 protostellar object L1157 has been previously reported as evidence for an accretion shock  \citep{Velusamy_2002}.

\citet{Ginsburg2023ApJ} took a sample of 23 candidate high-mass young stellar objects (HMYSO) in 17 high-mass star-forming regions found five new detections of \ce{H2O}, NaCl, KCl, PN, and SiS. While these detections had disk-like kinematic structures, SrcI is the only HMYSO that has been confirmed to have these molecules present in its disk. These results indicate that NaCl emission features are likely common in HMYSO disks.

% With 5 detections out of 23 HMYSOs, it may be that all of the detected sources have recently suffered severe perturbations.    Most massive stars are multiple, with large excess of close-in $<$ 10 AU companions).   If these are capture formed by dynamic interactions, 5/23 may not be an unreasonable fraction  which have stars orbiting in or near their disks which stir them up, creating shocks. The absence of explosive signatures in most may be a consequence of the short duration of explosive signatures (thousand of years at most) compared to the accretion time scale of a massive stars ($> \sim 10^{5}$ years).

\section{Summary}

1. We present ALMA observations of SiO, SiS, \ce{H2O}, NaCl, and SO at $\sim$30 to 50 mas resolution. These images map the molecular outflow and disk of Orion Source I (SrcI) on $\sim$12 to 20 AU scales.

2.  The SiO and \ce{^29SiO} emission map a rotating, expanding outflow. The emission is consistent with a magneto-centrifugal outflow from the SrcI disk. The rotational velocity decreases with distance from the disk.

3. The SiS maps a turbulent boundary layer in the outflow. We argue that the SiS emission is created and destroyed in a shocked region between the outflow and the ambient medium as SrcI moves through it. This process would dissipate the angular momentum held by the rotating molecular outflow into the surrounding medium.

4. The vibrationally excited water line (E$_{u}$ = 3461.9 K) traces the rotation of the SrcI disk and also traces the SiO outflow and SiS emission  close to the disk, suggesting that shocks may play a role in the  \ce{H2O} emission.
The origin of the vibrationally excited water line in the gas phase is not clear. The \ce{H2O} emission extends deeper and likely formed in the disk, but icy grain mantles could not persist in the hot disk. Hydrated minerals are a possible source of \ce{H2O} in the hot disk.

5. The NaCl emission closely traces the disk. The emission follows a linear velocity gradient along the disk major axis. At both ends of the disk, the position angle of the NaCl emission is consistent with a twist, or warp. Bright hot spots in NaCl emission may be excited by accretion shocks from material falling onto the disk. The motion of SrcI forces the external material to pass between the disk and the outflow, which may result in some material accreting onto the disk, creating shocks which excite the NaCl emission close to the disk surface.

6. SO traces a bow-shock at the leading edge of the SrcI disk and a turbulent plume of material extending from the trailing edge of the disk. This trailing material may accrete onto the disk. These features are created by the movement of SrcI through the medium and Bondi-Hoyle-Lyttleton accretion. 

7. If the accretion onto the circum-binary disk is the Bondi-Hoyle-Lyttleton rate $\sim$10$^{-7}$ $M_\odot$/yr and accretion from the disk onto the star is $\sim$10$^{-3}$ $M_\odot$/yr to maintain the observed luminosity of $\sim$10$^4$ $L_\odot$, then the disk is being depleted on a timescale 
% of $M_{disk}$/$\dot{M}$. \citet{Plambeck2016} estimated M$_{disk}$ $\sim$ 0.02 to 0.2$M_\odot$, giving a disk depletion 
time of 200 to 2000 yr.
%for $\dot{M}$$\sim$10$^{-3}$ $M_\odot/yr$.

8. These data suggest that the BN - SrcI interaction, and the proper motion of SrcI may have a significant effect on the structure and evolution of SrcI and its molecular outflow.
\facilities{ALMA, VLA}

\software{Miriad \citep{Sault1995}}
\begin{acknowledgements}
    
We thank the referee for a careful review and detailed suggestions to improve this paper.

The National Radio Astronomy Observatory is a facility of the National Science Foundation operated under cooperative agreement by Associated Universities, Inc. This paper makes use of the following ALMA data: 
2017.1.00497.S, 2016.1.00165.S, 2016.1.00165.S .
ALMA is a partnership of ESO (representing its member states), NSF (USA) and NINS (Japan), together with NRC (Canada), MOST and ASIAA (Taiwan), and KASI (Republic of Korea), in cooperation with the Republic of Chile. The Joint ALMA Observatory is operated by ESO, AUI/NRAO and NAOJ.
T.H. is financially supported by the MEXT/JSPS KAKENHI grant Nos.
17K05398, 18H05222, and 20H05845
\end{acknowledgements}

% --- table of observations --- %
\begin{deluxetable*}{cccccc}
\tabletypesize{\small}
\tablecaption{Observations}
\tablecolumns{6}
\tablenum{1}
\tablehead{
\colhead{freq} &
\colhead{project code} &
\colhead{date} &
\colhead{time} &
\colhead{synth beam}  &
\colhead{baseline} \\
\colhead{(GHz)} & & & 
\colhead{(min)} &
\colhead{(milliarcsec)} &
\colhead{(meters)}
}
\startdata
 %43  &  VLA/18A-136     &  2018-03-06  &  291  &   39$\times$34 at PA 1  & 500 - 36600  \\
 99  &  2017.1.00497.S	&  2017-10-12  &  158  &   45$\times$36 at PA 47  & 40 - 16200 \\
 %216-220  & 2013.1.005446.S  & 2014-12 to 2015-04  &   15    &   1500$\times$930 at PA -8 & 14 - 330 \\
224  &  2016.1.00165.S	&  2017-09-19  &   44  &   39$\times$19 at PA 66  & 40 - 10500 \\
340  &	2016.1.00165.S  &  2017-11-08  &   45  &   26$\times$11 at PA 58  & 90 - 12900 \\
%350  &	2012.1.00123.S  &  2014-07-26  &   24  &   276$\times$260 at PA 85 & 30 - 730 \\
\enddata
\end{deluxetable*}

\vspace{5mm}\

% --- table of observations --- %
\begin{deluxetable*}{CCCC}
\tabletypesize{\small}
\tablecaption{Spectral Lines}
\tablecolumns{4}
\tablenum{2}
\tablehead{
\colhead{Molecule} &
\colhead{Transition} &
\colhead{Frequency} &
\colhead{Upper Energy} \\
& &
\colhead{(GHz)} 
& 
\colhead{(K)} 
}
\startdata
 %43  &  VLA/18A-136     &  2018-03-06  &  291  &   39$\times$34 at PA 1  & 500 - 36600  \\
 ^{28}\textrm{SiO} & v = 0, J = 5 - 4 & 217.105 & 14.5 \\
 ^{29}\textrm{SiO} & v = 0, J = 5 - 4 & 214.385 & 30.9 \\
 \textrm{SiS} & v = 0, J = 12 - 11 & 217.818 & 68.0 \\
 % 344.31 GHz SO emission
 \textrm{H}_{2}\textrm{O} & 5( 5, 0), v_{2}=1 - 6( 4, 3), v_{2}=1 & 232.687 & 3461.9 \\
 %  217.98 GHz NaCl $v=2$ $J=17-16$ 
 % 23 35 2 17 16 1128.4 0.00490 217.98023 D
 \textrm{NaCl} & v = 1, J = 18 - 17 & 232.510 & 625.7 \\
 \textrm{NaCl} & v = 2, J = 17 - 16 & 217.980 & 1128.4 \\
% Sulfur Monoxide	344.3106120000001, 344.310612	8( 8)- 7( 7)	-1.8699	10.93	70.95734	87.48155
\textrm{SO} & ^{3}\Sigma v = 0, J = 5 - 4 & 215.221 & 44.1 \\
\textrm{SO} & ^{3}\Sigma v = 0, J = 8 - 7 & 344.311 & 87.5 \\
\enddata
\end{deluxetable*}

\vspace{5mm}

\bibliographystyle{aasjournal}
\bibliography{SrcI.bib, brett_sis}

\begin{thebibliography}{}
\expandafter\ifx\csname natexlab\endcsname\relax\def\natexlab#1{#1}\fi
\providecommand{\url}[1]{\href{#1}{#1}}

\bibitem[{{Bally} {et~al.}(2017){Bally}, {Ginsburg}, {Arce}, {Eisner},
  {Youngblood}, {Zapata}, \& {Zinnecker}}]{Bally2017}
{Bally}, J., {Ginsburg}, A., {Arce}, H., {et~al.} 2017, \apj, 837, 60

\bibitem[{{Bally} {et~al.}(2020){Bally}, {Ginsburg}, {Forbrich}, \&
  {Vargas-Gonz{\'a}lez}}]{Bally2020}
{Bally}, J., {Ginsburg}, A., {Forbrich}, J., \& {Vargas-Gonz{\'a}lez}, J. 2020,
  \apj, 889, 178

\bibitem[{{Bonnell}(2008)}]{Bonnell_2008}
{Bonnell}, I.~A. 2008, in Astronomical Society of the Pacific Conference
  Series, Vol. 390, Pathways Through an Eclectic Universe, ed. J.~H. {Knapen},
  T.~J. {Mahoney}, \& A.~{Vazdekis}, 26

\bibitem[{Bonnell {et~al.}(1998)Bonnell, Bate, \& Zinnecker}]{Bonnell_1998}
Bonnell, I.~A., Bate, M.~R., \& Zinnecker, H. 1998, Monthly Notices of the
  Royal Astronomical Society, 298, 93.
\newblock \url{https://doi.org/10.1046/j.1365-8711.1998.01590.x}

\bibitem[{Burkhardt {et~al.}(2019)Burkhardt, Shingledecker, Gal, McGuire,
  Remijan, \& Herbst}]{Burkhardt_2019}
Burkhardt, A.~M., Shingledecker, C.~N., Gal, R.~L., {et~al.} 2019, The
  Astrophysical Journal, 881, 32.
\newblock \url{https://dx.doi.org/10.3847/1538-4357/ab2be8}

\bibitem[{Campanha {et~al.}(2022)Campanha, Mendoza, Silva, Velloso, Carvajal,
  Wakelam, \& Galv{\~a}o}]{Campanha:2022:369}
Campanha, D.~R., Mendoza, E., Silva, M.~X., {et~al.} 2022, Monthly Notices of
  the Royal Astronomical Society, 515, 369

\bibitem[{{Caratti o Garatti} {et~al.}(2017){Caratti o Garatti}, {Stecklum},
  {Garcia Lopez}, {Eisl{\"o}ffel}, {Ray}, {Sanna}, {Cesaroni}, {Walmsley},
  {Oudmaijer}, {de Wit}, {Moscadelli}, {Greiner}, {Krabbe}, {Fischer}, {Klein},
  \& {Iba{\~n}ez}}]{Caretti_o_Garatti_2017}
{Caratti o Garatti}, A., {Stecklum}, B., {Garcia Lopez}, R., {et~al.} 2017,
  Nature Physics, 13, 276

\bibitem[{{Charnley}(1997)}]{Charnley_1997}
{Charnley}, S.~B. 1997, \apj, 481, 396

\bibitem[{{Commer{\c{c}}on} {et~al.}(2022){Commer{\c{c}}on}, {Gonz{\'a}lez},
  {Mignon-Risse}, {Hennebelle}, \& {Vaytet}}]{Commercon2022}
{Commer{\c{c}}on}, B., {Gonz{\'a}lez}, M., {Mignon-Risse}, R., {Hennebelle},
  P., \& {Vaytet}, N. 2022, \aap, 658, A52

\bibitem[{Doddipatla {et~al.}(2021)Doddipatla, He, Goettl, Kaiser, Galv{\~a}o,
  \& Millar}]{Doddipatla:2021:eabg7003}
Doddipatla, S., He, C., Goettl, S.~J., {et~al.} 2021, Science Advances, 7,
  eabg7003

\bibitem[{{Dzib} {et~al.}(2017){Dzib}, {Loinard}, {Rodr{\'\i}guez},
  {G{\'o}mez}, {Forbrich}, {Menten}, {Kounkel}, {Mioduszewski}, {Hartmann},
  {Tobin}, \& {Rivera}}]{Dzib_2017}
{Dzib}, S.~A., {Loinard}, L., {Rodr{\'\i}guez}, L.~F., {et~al.} 2017, \apj,
  834, 139

\bibitem[{Edgar(2004)}]{Edgar_2004}
Edgar, R. 2004, New Astronomy Reviews, 48, 843–859.
\newblock \url{http://dx.doi.org/10.1016/j.newar.2004.06.001}

\bibitem[{{Ginsburg} {et~al.}(2018){Ginsburg}, {Bally}, {Goddi}, {Plambeck}, \&
  {Wright}}]{Ginsburg2018}
{Ginsburg}, A., {Bally}, J., {Goddi}, C., {Plambeck}, R., \& {Wright}, M. 2018,
  \apj, 860, 119

\bibitem[{Ginsburg {et~al.}(2019)Ginsburg, McGuire, Plambeck, Bally, Goddi, \&
  Wright}]{Ginsburg2019}
Ginsburg, A., McGuire, B., Plambeck, R., {et~al.} 2019, The Astrophysical
  Journal, 872, 54.
\newblock \url{https://doi.org/10.3847%2F1538-4357%2Faafb71}

\bibitem[{{Ginsburg} {et~al.}(2023){Ginsburg}, {McGuire}, {Sanhueza}, {Olguin},
  {Maud}, {Tanaka}, {Zhang}, {Beuther}, \& {Indriolo}}]{Ginsburg2023ApJ}
{Ginsburg}, A., {McGuire}, B.~A., {Sanhueza}, P., {et~al.} 2023, \apj, 942, 66

\bibitem[{{Goddi} {et~al.}(2009){Goddi}, {Greenhill}, {Chandler}, {Humphreys},
  {Matthews}, \& {Gray}}]{Goddi2009}
{Goddi}, C., {Greenhill}, L.~J., {Chandler}, C.~J., {et~al.} 2009, \apj, 698,
  1165

\bibitem[{{Goddi} {et~al.}(2011{\natexlab{a}}){Goddi}, {Greenhill},
  {Humphreys}, {Chandler}, \& {Matthews}}]{Goddi2011b}
{Goddi}, C., {Greenhill}, L.~J., {Humphreys}, E.~M.~L., {Chandler}, C.~J., \&
  {Matthews}, L.~D. 2011{\natexlab{a}}, \apjl, 739, L13

\bibitem[{{Goddi} {et~al.}(2011{\natexlab{b}}){Goddi}, {Humphreys},
  {Greenhill}, {Chandler}, \& {Matthews}}]{Goddi2011}
{Goddi}, C., {Humphreys}, E.~M.~L., {Greenhill}, L.~J., {Chandler}, C.~J., \&
  {Matthews}, L.~D. 2011{\natexlab{b}}, \apj, 728, 15

\bibitem[{{Greenhill} {et~al.}(2013){Greenhill}, {Goddi}, {Chandler},
  {Matthews}, \& {Humphreys}}]{Greenhill2013}
{Greenhill}, L.~J., {Goddi}, C., {Chandler}, C.~J., {Matthews}, L.~D., \&
  {Humphreys}, E.~M.~L. 2013, \apjl, 770, L32

\bibitem[{{Hirota} {et~al.}(2014){Hirota}, {Kim}, {Kurono}, \&
  {Honma}}]{Hirota2014}
{Hirota}, T., {Kim}, M.~K., {Kurono}, Y., \& {Honma}, M. 2014, \apjl, 782, L28

\bibitem[{{Hirota} {et~al.}(2017){Hirota}, {Machida}, {Matsushita}, {Motogi},
  {Matsumoto}, {Kim}, {Burns}, \& {Honma}}]{Hirota2017}
{Hirota}, T., {Machida}, M.~N., {Matsushita}, Y., {et~al.} 2017, Nature
  Astronomy, 1, 0146

\bibitem[{{Hirota} {et~al.}(2020){Hirota}, {Plambeck}, {Wright}, {Machida},
  {Matsushita}, {Motogi}, {Kim}, {Burns}, \& {Honma}}]{Hirota2020}
{Hirota}, T., {Plambeck}, R.~L., {Wright}, M. C.~H., {et~al.} 2020, \apj, 896,
  157

\bibitem[{Hosokawa {et~al.}(2010)Hosokawa, Yorke, \& Omukai}]{Hosokawa2010}
Hosokawa, T., Yorke, H.~W., \& Omukai, K. 2010, The Astrophysical Journal, 721,
  478.
\newblock \url{https://dx.doi.org/10.1088/0004-637X/721/1/478}

\bibitem[{Indriolo {et~al.}(2013)Indriolo, Neufeld, Seifahrt, \&
  Richter}]{Indriolo_2013}
Indriolo, N., Neufeld, D.~A., Seifahrt, A., \& Richter, M.~J. 2013, The
  Astrophysical Journal, 776, 8.
\newblock \url{https://dx.doi.org/10.1088/0004-637X/776/1/8}

\bibitem[{{Kim} {et~al.}(2008){Kim}, {Hirota}, {Honma}, {Kobayashi},
  {Bushimata}, {Choi}, {Imai}, {Iwadate}, {Jike}, {Kameno}, {Kameya},
  {Kamohara}, {Kan-Ya}, {Kawaguchi}, {Kuji}, {Kurayama}, {Manabe}, {Matsui},
  {Matsumoto}, {Miyaji}, {Nagayama}, {Nakagawa}, {Oh}, {Omodaka}, {Oyama},
  {Sakai}, {Sasao}, {Sato}, {Sato}, {Shibata}, {Tamura}, \&
  {Yamashita}}]{Kim2008}
{Kim}, M.~K., {Hirota}, T., {Honma}, M., {et~al.} 2008, \pasj, 60, 991

\bibitem[{{Kounkel} {et~al.}(2018){Kounkel}, {Covey}, {Su{\'a}rez},
  {Hernandez}, {Stassun}, {Jaehnig}, {Feigelson}, {Pe{\~n}a Ram{\'\i}rez},
  {Roman-Lopes}, {Da Rio}, {Stringfellow}, {Kim}, {Borissova},
  {Fern{\'a}ndez-Trincado}, {Burgasser}, {Garc{\'\i}a-Hern{\'a}ndez}, {Zamora},
  {Pan}, \& {Nitschelm}}]{Kounkel2018}
{Kounkel}, M., {Covey}, K., {Su{\'a}rez}, Genaro a
  nd~{Rom{\'a}n-Z{\'u}{\~n}iga}, C., {et~al.} 2018, \aj, 156, 84

\bibitem[{{Krumholz} {et~al.}(2009){Krumholz}, {Klein}, {McKee}, {Offner}, \&
  {Cunningham}}]{Krumholz2009}
{Krumholz}, M.~R., {Klein}, R.~I., {McKee}, C.~F., {Offner}, S. S.~R., \&
  {Cunningham}, A.~J. 2009, Science, 323, 754

\bibitem[{Kuiper {et~al.}(2011)Kuiper, Klahr, Beuther, \&
  Henning}]{Kuiper_2011}
Kuiper, R., Klahr, H., Beuther, H., \& Henning, T. 2011, The Astrophysical
  Journal, 732, 20.
\newblock \url{http://dx.doi.org/10.1088/0004-637X/732/1/20}

\bibitem[{{Lenzuni} {et~al.}(1995){Lenzuni}, {Gail}, \&
  {Henning}}]{Lenzuni1995}
{Lenzuni}, P., {Gail}, H.-P., \& {Henning}, T. 1995, \apj, 447, 848

\bibitem[{Li {et~al.}(2015)Li, Wang, Zhu, Zhang, \& Li}]{Li_2015}
Li, J., Wang, J., Zhu, Q., Zhang, J., \& Li, D. 2015, The Astrophysical
  Journal, 802, 40.
\newblock \url{https://dx.doi.org/10.1088/0004-637X/802/1/40}

\bibitem[{{Luhman} {et~al.}(2017){Luhman}, {Robberto}, {Tan}, {Andersen},
  {Giulia Ubeira Gabellini}, {Manara}, {Platais}, \& {Ubeda}}]{Luhman_2017}
{Luhman}, K.~L., {Robberto}, M., {Tan}, J.~C., {et~al.} 2017, \apjl, 838, L3

\bibitem[{López-Vázquez {et~al.}(2020)López-Vázquez, Zapata, Lizano, \&
  Cantó}]{Lopez-Vazquez2020}
López-Vázquez, J.~A., Zapata, L.~A., Lizano, S., \& Cantó, J. 2020, The
  Astrophysical Journal, 904, 158.
\newblock \url{http://dx.doi.org/10.3847/1538-4357/abbe24}

\bibitem[{Machida {et~al.}(2019)Machida, Hirano, \& Kitta}]{Machida_2019}
Machida, M.~N., Hirano, S., \& Kitta, H. 2019, Monthly Notices of the Royal
  Astronomical Society, doi:10.1093/mnras/stz3159.
\newblock \url{http://dx.doi.org/10.1093/mnras/stz3159}

\bibitem[{{Matthews} {et~al.}(2010){Matthews}, {Greenhill}, {Goddi}, {Chand
  ler}, {Humphreys}, \& {Kunz}}]{Matthews2010}
{Matthews}, L.~D., {Greenhill}, L.~J., {Goddi}, C., {et~al.} 2010, \apj, 708,
  80

\bibitem[{{Maud} {et~al.}(2019){Maud}, {Cesaroni}, {Kumar}, {Rivilla},
  {Ginsburg}, {Klaassen}, {Harsono}, {S{\'a}nchez-Monge}, {Ahmadi}, {Allen},
  {Beltr{\'a}n}, {Beuther}, {Galv{\'a}n-Madrid}, {Goddi}, {Hoare},
  {Hogerheijde}, {Johnston}, {Kuiper}, {Moscadelli}, {Peters}, {Testi}, {van
  der Tak}, \& {de Wit}}]{Maud_2019}
{Maud}, L.~T., {Cesaroni}, R., {Kumar}, M.~S.~N., {et~al.} 2019, \aap, 627, L6

\bibitem[{{Menten} {et~al.}(2007){Menten}, {Reid}, {Forbrich}, \&
  {Brunthaler}}]{Menten2007}
{Menten}, K.~M., {Reid}, M.~J., {Forbrich}, J., \& {Brunthaler}, A. 2007, \aap,
  474, 515

\bibitem[{{Minh} {et~al.}(1990){Minh}, {Ziurys}, {Irvine}, \&
  {McGonagle}}]{Minh_1990}
{Minh}, Y.~C., {Ziurys}, L.~M., {Irvine}, W.~M., \& {McGonagle}, D. 1990, \apj,
  360, 136

\bibitem[{Moeckel \& Throop(2009)}]{Moeckel_2009}
Moeckel, N., \& Throop, H.~B. 2009, The Astrophysical Journal, 707, 268.
\newblock \url{https://dx.doi.org/10.1088/0004-637X/707/1/268}

\bibitem[{Neufeld {et~al.}(2012)Neufeld, Falgarone, Gerin, Godard, Herbst,
  Pineau Des~For{\^e}ts, Vasyunin, G{\"u}sten, Wiesemeyer, \&
  Ricken}]{Neufeld:2012:L6}
Neufeld, D.~A., Falgarone, E., Gerin, M., {et~al.} 2012, Astronomy \&
  Astrophysics, 542, L6

\bibitem[{{Niederhofer} {et~al.}(2012){Niederhofer}, {Humphreys}, \&
  {Goddi}}]{Niederhofer2012}
{Niederhofer}, F., {Humphreys}, E.~M.~L., \& {Goddi}, C. 2012, \aap, 548, A69

\bibitem[{Paiva {et~al.}(2020)Paiva, Lefloch, \& Galv{\~a}o}]{Paiva:2020:299}
Paiva, M. A.~M., Lefloch, B., \& Galv{\~a}o, B. R.~L. 2020, Monthly Notices of
  the Royal Astronomical Society, 493, 299

\bibitem[{{Pillich} {et~al.}(2021){Pillich}, {Bogdan}, {Landers}, {Wurm}, \&
  {Wende}}]{Pilloch_2021}
{Pillich}, C., {Bogdan}, T., {Landers}, J., {Wurm}, G., \& {Wende}, H. 2021,
  \aap, 652, A106

\bibitem[{{Pineau des Forets} {et~al.}(1993){Pineau des Forets}, {Roueff},
  {Schilke}, \& {Flower}}]{Pineau1993}
{Pineau des Forets}, G., {Roueff}, E., {Schilke}, P., \& {Flower}, D.~R. 1993,
  \mnras, 262, 915

\bibitem[{{Plambeck} \& {Wright}(2016)}]{Plambeck2016}
{Plambeck}, R.~L., \& {Wright}, M.~C.~H. 2016, \apj, 833, 219

\bibitem[{{Plambeck} {et~al.}(2009){Plambeck}, {Wright}, {Friedel}, {Widicus
  Weaver}, {Bolatto}, {Pound}, {Woody}, {Lamb}, \& {Scott}}]{Plambeck2009}
{Plambeck}, R.~L., {Wright}, M.~C.~H., {Friedel}, D.~N., {et~al.} 2009, \apjl,
  704, L25

\bibitem[{Pudritz \& Ray(2019)}]{Pudritz_2019}
Pudritz, R.~E., \& Ray, T.~P. 2019, Frontiers in Astronomy and Space Sciences,
  6, doi:10.3389/fspas.2019.00054.
\newblock \url{https://www.frontiersin.org/articles/10.3389/fspas.2019.00054}

\bibitem[{{Reid} {et~al.}(2007){Reid}, {Menten}, {Greenhill}, \&
  {Chandler}}]{Reid2007}
{Reid}, M.~J., {Menten}, K.~M., {Greenhill}, L.~J., \& {Chandler}, C.~J. 2007,
  \apj, 664, 950

\bibitem[{{Rodr{\'{\i}}guez} {et~al.}(2017){Rodr{\'{\i}}guez}, {Dzib},
  {Loinard}, {Zapata}, {G{\'o}mez}, {Menten}, \& {Lizano}}]{Rodgriguez2017}
{Rodr{\'{\i}}guez}, L.~F., {Dzib}, S.~A., {Loinard}, L., {et~al.} 2017, \apj,
  834, 140

\bibitem[{Rodríguez {et~al.}(2017)Rodríguez, Dzib, Loinard, Zapata, Gómez,
  Menten, \& Lizano}]{Rodriguez2017}
Rodríguez, L.~F., Dzib, S.~A., Loinard, L., {et~al.} 2017, The Astrophysical
  Journal, 834, 140.
\newblock \url{https://dx.doi.org/10.3847/1538-4357/834/2/140}

\bibitem[{Rodríguez {et~al.}(2020)Rodríguez, Dzib, Zapata, Lizano, Loinard,
  Menten, \& Gómez}]{Rodriguez2020}
Rodríguez, L.~F., Dzib, S.~A., Zapata, L., {et~al.} 2020, The Astrophysical
  Journal, 892, 82.
\newblock \url{https://dx.doi.org/10.3847/1538-4357/ab7816}

\bibitem[{Rosi {et~al.}(2018)Rosi, Mancini, Skouteris, Ceccarelli,
  Faginas~Lago, Podio, Codella, Lefloch, \& Balucani}]{Rosi:2018:87}
Rosi, M., Mancini, L., Skouteris, D., {et~al.} 2018, Chemical Physics Letters,
  695, 87

\bibitem[{{Sault} {et~al.}(1995){Sault}, {Teuben}, \& {Wright}}]{Sault1995}
{Sault}, R.~J., {Teuben}, P.~J., \& {Wright}, M.~C.~H. 1995, in Astronomical
  Society of the Pacific Conference Series, Vol.~77, Astronomical Data Analysis
  Software and Systems IV, ed. R.~A. {Shaw}, H.~E. {Payne}, \& J.~J.~E.
  {Hayes}, 433

\bibitem[{Schilke {et~al.}(2001)Schilke, Benford, Hunter, Lis, \&
  Phillips}]{Schilke:2001:281}
Schilke, P., Benford, D.~J., Hunter, T.~R., Lis, D.~C., \& Phillips, T.~G.
  2001, The Astrophysical Journal Supplement Series, 132, 281

\bibitem[{{Schilke} {et~al.}(1997){Schilke}, {Walmsley}, {Pineau des Forets},
  \& {Flower}}]{Schilke1997}
{Schilke}, P., {Walmsley}, C.~M., {Pineau des Forets}, G., \& {Flower}, D.~R.
  1997, \aap, 321, 293

\bibitem[{Shu(1991)}]{shu1991physics}
Shu, F. 1991, The Physics of Astrophysics: Gas dynamics, Series of books in
  astronomy (University Science Books).
\newblock \url{https://books.google.com/books?id=50VYSc56URUC}

\bibitem[{{Shu} {et~al.}(2000){Shu}, {Najita}, {Shang}, \& {Li}}]{Shu2000}
{Shu}, F.~H., {Najita}, J.~R., {Shang}, H., \& {Li}, Z.-Y. 2000, Protostars and
  Planets IV, 789

\bibitem[{{Snell} {et~al.}(1984){Snell}, {Scoville}, {Sanders}, \&
  {Erickson}}]{Snell1984}
{Snell}, R.~L., {Scoville}, N.~Z., {Sanders}, D.~B., \& {Erickson}, N.~R. 1984,
  \apj, 284, 176

\bibitem[{{Tanaka} {et~al.}(2020){Tanaka}, {Zhang}, {Hirota}, {Sakai},
  {Motogi}, {Tomida}, {Tan}, {Rosero}, {Higuchi}, {Ohashi}, {Liu}, \&
  {Sugiyama}}]{Tanaka2020}
{Tanaka}, K. E.~I., {Zhang}, Y., {Hirota}, T., {et~al.} 2020, \apjl, 900, L2

\bibitem[{{Testi} {et~al.}(2010){Testi}, {Tan}, \& {Palla}}]{Testi2010}
{Testi}, L., {Tan}, J.~C., \& {Palla}, F. 2010, \aap, 522, A44

\bibitem[{{Ulrich}(1976)}]{Ulrich1976}
{Ulrich}, R.~K. 1976, \apj, 210, 377

\bibitem[{{van der Tak} {et~al.}(2006){van der Tak}, {Walmsley}, {Herpin}, \&
  {Ceccarelli}}]{van_der_Tak_2006}
{van der Tak}, F.~F.~S., {Walmsley}, C.~M., {Herpin}, F., \& {Ceccarelli}, C.
  2006, \aap, 447, 1011

\bibitem[{van Dishoeck \& Blake(1998)}]{van_Dishoeck_1998}
van Dishoeck, E.~F., \& Blake, G.~A. 1998, Annual Review of Astronomy and
  Astrophysics, 36, 317.
\newblock \url{https://doi.org/10.1146/annurev.astro.36.1.317}

\bibitem[{{van Gelder, M. L.} {et~al.}(2021){van Gelder, M. L.}, {Tabone, B.},
  {van Dishoeck, E. F.}, \& {Godard, B.}}]{vanGelder_2021}
{van Gelder, M. L.}, {Tabone, B.}, {van Dishoeck, E. F.}, \& {Godard, B.} 2021,
  A\&A, 653, A159.
\newblock \url{https://doi.org/10.1051/0004-6361/202141591}

\bibitem[{{Velusamy} {et~al.}(2002){Velusamy}, {Langer}, \&
  {Goldsmith}}]{Velusamy_2002}
{Velusamy}, T., {Langer}, W.~D., \& {Goldsmith}, P.~F. 2002, \apjl, 565, L43

\bibitem[{{Wakelam} {et~al.}(2005){Wakelam}, {Ceccarelli}, {Castets},
  {Lefloch}, {Loinard}, {Faure}, {Schneider}, \& {Benayoun}}]{Wakelam_2005}
{Wakelam}, V., {Ceccarelli}, C., {Castets}, A., {et~al.} 2005, \aap, 437, 149

\bibitem[{{Wakelam} {et~al.}(2011){Wakelam}, {Hersant}, \&
  {Herpin}}]{Wakelam_2011}
{Wakelam}, V., {Hersant}, F., \& {Herpin}, F. 2011, \aap, 529, A112

\bibitem[{Whitney {et~al.}(2013)Whitney, Robitaille, Bjorkman, Dong, Wolff,
  Wood, \& Honor}]{Whitney_2013}
Whitney, B.~A., Robitaille, T.~P., Bjorkman, J.~E., {et~al.} 2013, The
  Astrophysical Journal Supplement Series, 207, 30.
\newblock \url{http://dx.doi.org/10.1088/0067-0049/207/2/30}

\bibitem[{{Wright} {et~al.}(2020){Wright}, {Plambeck}, {Hirota}, {Ginsburg},
  {McGuire}, {Bally}, \& {Goddi}}]{Wright2020}
{Wright}, M., {Plambeck}, R., {Hirota}, T., {et~al.} 2020, \apj, 889, 155

\bibitem[{{Wright} {et~al.}(2022){Wright}, {Bally}, {Hirota}, {Miller},
  {Harding}, {Colleluori}, {Ginsburg}, {Goddi}, \& {McGuire}}]{Wright2022}
{Wright}, M., {Bally}, J., {Hirota}, T., {et~al.} 2022, \apj, 924, 107

\end{thebibliography}

%\input{table2.tex}

% FIGURE 1

% FIGURE 2

% FIGURE 3

% FIGURE 4

% FIGURE 5
% sio+sis.pdf

% FIGURE 6

% FIGURE 7

% FIGURE 8

% Figure 9

% Figure 10

% FIGURE 11

% Figure 12

% FIGURE 13

% FIGURE 14

 % FIGURE 15

\end{document}